\documentclass[A4]{article}
\usepackage[T1]{fontenc}
\usepackage[latin1]{inputenc}
\usepackage{amsmath, amssymb, amsthm}
\usepackage[english]{babel}
\usepackage{indentfirst}
\usepackage{booktabs}
\usepackage{array}
\usepackage{caption,appendix}
\usepackage{geometry}
\usepackage{float}
\usepackage{cancel}
\usepackage{bbold}
\usepackage{authblk}
\numberwithin{equation}{section}
\begin{document}
\author[1,2,3,4]{Salvatore Capozziello}
\author[5]{Maurizio Capriolo} 
\author[5]{Maria Transirico}
\affil[1]{\emph{Dipartimento di Fisica "E. Pancini", Universit\`a		   di Napoli {}``Federico II'', Compl. Univ. di
		   Monte S. Angelo, Edificio G, Via Cinthia, I-80126, Napoli, Italy, }}
		  \affil[2]{\emph{INFN Sezione  di Napoli, Compl. Univ. di
		   Monte S. Angelo, Edificio G, Via Cinthia, I-80126,  Napoli, Italy,}}
 \affil[3]{\emph{Gran Sasso Science Institute, Via F. Crispi 			   	   7, I-67100, L'Aquila, Italy,}}
 \affil[4]{\emph{ Tomsk State Pedagogical University, ul. Kievskaya, 60, 634061
Tomsk, Russia, }}
\affil[5]{\emph{Dipartimento di Matematica Universit\`a di Salerno, via Giovanni Paolo II, 132, Fisciano, SA I-84084, Italy.} }
\date{\today}
\title{\textbf{The gravitational energy-momentum  pseudotensor:  the cases of $f(R)$ and $f(T)$ gravity}}
\maketitle
\begin{abstract}
 We derive the gravitational energy-momentum  pseudotensor $ \tau^{\sigma}_ {\phantom {\sigma} \lambda} $ in metric $ f\left (R \right) $ gravity and in  
 teleparallel $ f\left (T\right) $ gravity. In the first case,   $R$ is the Ricci curvature scalar for a torsionless Levi-Civita connection; in the second case,  $T$ is the curvature-free torsion scalar derived by  tetrads and Weitzenb\"ock connection.    For both classes of  theories the continuity equations are  obtained in  presence of matter.   $ f \left (R \right) $ and  $ f \left (T \right) $ are non-equivalent  but differ for a quantity 
 $ \omega \left (T, B \right) $ containing  the torsion scalar $T$ and a boundary term $ B $.  It is possible to obtain  the field equations for  
 $ \omega \left (T, B \right) $ and the related  gravitational energy-momentum pseudotensor $ \tau^{\sigma}_{\phantom {\sigma}\lambda} \vert \omega $. Finally we  show that, thanks to this further pseudotensor, it is possible to pass from  $ f \left (R \right) $ to  $ f \left ( T \right) $ and viceversa through a simple relation between  gravitational  pseudotensors.
\end{abstract}

\section{Introduction}
The issue to extend General Relativity (GR) is related to the necessity to obtain a comprehensive picture of gravitational interaction from UV to IR scales ranging from  quantum gravity, to  inflation,  large scale structure  and  today observed accelerated behavior of the Hubble flow. 
The standard approach is introducing dark components (dark energy and dark matter) and scalar fields (e.g. the inflaton) within the framework of Einstein's  GR. The question is today completely open because no final evidence for exotic matter and fields has  been reported. 

An alternative way to proceed  is to modify Einstein's theory  by constructing classes of theories where GR is a particular case. This is the main issue by which  Extended Theories of Gravitation \cite{CL} are born\footnote{We like to call these theories "Extended Gravity" instead of "Alternative Gravity" because GR is a so "well-posed",  "well-formulated", and experimentally probed  theory that, at present state of art,  it does not need alternatives but just extensions at the various scales and comparison with new physical problems. }.  The simplest family is the so called  $ f \left (R \right) $ gravity  that is a generalization  of the Einstein-Hilbert action which is not simply  linear in the Ricci scalar $R$.  The straightforward approach is adopting  a Levi Civita metric connection  without  torsion. In this scheme,   gravity is interpreted in terms of the curvature of  a spacetime manifold,  the free fall of bodies happens  along  geodesics and the dynamical  variables are the components of the metric tensor.  An alternative formulation of Einstein's gravity  is the Teleparallel Equivalent of General Relativity (TEGR) \cite{AE,UC} based on the  parallelism of tetrads (vierbiens), an orthonormal basis in the Minkowski tangent space, at any point, of a given spacetime manifold. The geometric structure results with   zero curvature and  non-zero torsion. The   connection is   the Weitzenb\"ock  connection \cite{WIT, APTG, Laur, MAL}. In teleparallel gravity, the dynamical  variables are the tetrads, the gravity returns to be interpreted as a force and the bodies follow the motion governed by the equations of the gravitational force produced by  torsion: therefore it is the torsion to produce the gravity in TEGR;  the action is equivalent to the Einstein-Hilbert one but it is  built with the  torsion scalar $T$ that replaces the  curvature scalar $ R $.  

 TEGR can be generalized in the sense of  $ f \left (R \right) $, that is   $f \left (T \right) $ models  can be considered where $f(T)$ is a generic function of the torsion scalar $T$  \cite {CCLS, FFPL}. However, while  $ f \left (R \right) $ gravity is  invariant for local Lorentz transformations and shows fourth-order field equations in the metric formalism,  $ f \left (T \right) $ gravity is not invariant for local Lorentz transformations  of  vierbein fields and has second-order field  equations. The lack of local Lorentz invariance is due to the choice of the Weitzenb\"ock connection which implies the cancellation of the spin connection which leads to  a particular tetrad  called {\it pure tetrad}. We can recover this invariance by choosing a spin connection which  is different from zero. Furthermore covariant  $f(T)$ gravity has ben widely discussed in literature   \cite{BSB, CRTEL, KIB, FF, SLB,Emmanuel2}. In addition, while  $f \left (R \right) $ gravity is conformally  equivalent to Einstein's theory plus a scalar field,  $ f \left (T \right) $ gravity, in general,  is  not conformally  equivalent to TEGR plus  a scalar field \cite{CCLS,FFPL, Bamba, Emmanuel1, Martin1,Martin2}. It is also possible to interpret the telepallel theories as gauge theories for the group of translations that acts on the tangent fiber bundle at each point of the manifold \cite{APTG, Laur,Cho}.  Finally $ f \left (R \right) $ and $ f \left (T \right) $ theories can be generalized considering   derivative terms of both curvature and torsion, or assuming further geometric invariants  \cite{OS, KS,Kostas,felixgauss}. 
 
 The aim of this work is to discuss  the gravitational energy-momentum pseudotensor  \cite {LLF, MWH} in  $ f \left (R \right) $ and $ f \left (T \right) $ gravity,  to write the related  field equations, and to derive the  conservation laws in  presence of matter along the scheme discussed  in  \cite{CCT} (see also \cite{vagenas,abedi1}).  The features of gravitational energy-momentum pseudotensor  point out intrinsic differences between  metric and teleparallel formulations of theories of gravity. The first result  is that 
 the two  theories  differ for a term $ \omega \left (T, B \right) $ containing the  torsion scalar $T$  and a boundary contribution $B$  by which the  local Lorentz  invariance can be restored (see also \cite{BBW,BC}).  The gravitational energy-momentum pseudotensor, related to $\omega (T,B)$, allows to pass  from  $ f \left (R \right) $ to  $f \left (T \right) $ and viceversa.
 
The paper is organized as follows. In Sec.\ref {GS},  we give a short summary on geometrical setting considering, in particular, the   teleparallelism.  
In Sec.\ref{GEMTR}, we derive   the Euler-Lagrange equations and  the gravitational energy-momentum pseudotensor for $f(R)$ gravity by applying the Noether theorem to a particular  continuous one-parameter group   of diffeomorphisms representing  rigid translations.  Furthermore, we obtain  the continuity equation  in the presence of matter. In Sec.\ref {GEMPT}, we write the field equations for $f(T)$  gravity  and obtain the gravitational energy-momentum pseudotensor. As above,  we derive  the continuity equation  in presence of matter.  Sec.\ref{GEMPB} is devoted to the gravitational pseudotensor for  the $ \omega \left (T, B \right) $ term.
Finally,  in Sec.\ref {JUMPTB}, through the three obtained pseudotensors, we discuss the relation between  the two  theories of gravitation. In Sec.\ref{CONC}, conclusions are drawn.  In Appendix \ref{A}, we report  a list of useful formulas and, in Appendix \ref{B}, a list of  variations used along the paper.

\section{Geometrical setting}\label{GS}
Spacetime ${\cal M}$ is a   pseudo-Riemannian manifold  $\left\{{\cal M},\textnormal{\bf g}\right\}$  where a metric tensor  {\bf g} is defined.  It is a  non-degenerate symmetric tensor field  $\left(0,2\right)$,   non-positive defined with signature $(1,3)=(+,-,-,-)$. Furthermore  ${\cal M}$ is a Hausdorff topological manifold,  locally  omeomorphic  to a set in  ${\mathbb R}^4$, equipped with a differential structure  that is a maximal atlas  covering it. A \emph{fiber bundle} on ${\cal M}$ with standard fiber $\cal S$, is a  quadruple  $\left \{{\cal E}, \pi, {\cal M}, {\cal S} \right\}$, where ${\cal E, M, S}$ are  differential manifolds and   $\pi : {\cal E \rightarrow M}$,  is a surjective differential application, i.e. a  projection, locally trivial   in $\cal E$ that is  $\forall p\in {\cal M}$ exists a set  $ U$  in $\cal M$  and  a  diffeomorphism  $\chi:\pi^{-1}\left(U\right)\rightarrow U\times \cal S$  so that  $\pi_{1}\circ\chi =\pi$ where  $\pi_{1}$ is  the  projection along  the first   coordinate. Let $G$ be a Lie group. A \emph{G-bundle structure} on the fiber bundle consists of a left action $\theta : G\times \cal S$ and of a fiber bundle atlas $(U_{\alpha},\psi_{\alpha})$ whose transition functions $\psi_{\alpha\beta}$ act on $\cal S$ via the $G$-action.  A fiber bundle with a $G$-bundle structure is called a \emph{$G$-bundle}. A \emph{principal (fiber) bundle} $\left(P, \pi, {\cal M}, G\right)$ is a $G$-bundle with typical fiber which is a Lie group $G$, where the left action  on $G$ is just the left translation that is a fiber bundle whose fiber coincides with the structure group. The \emph{frame bundle } is the disjoint union, for each $p\in \cal M$,  of all  linear bases of the tangent space $T_{p}$ in $p$ (fiber in $p$) which is a principal fiber bundle with a $GL\left(m,\mathbb R\right)$ group structure  acting on the  fiber in $p$. A connection $\Gamma$ is a structure defined on a principal bundle $P$. It is given on  a \emph{horizontal subspace} $H_{u}$ of $T_{u}\left(P\right)$, the tangent space of $P$ in $u\in P$, for each $u\in P$ such that $T_{u}\left(P\right)=V_{u}\oplus H_{u}$, namely $ T\left(P\right)=V\oplus H$  and $u\rightarrow V_{u}$ is differentiable and invariant under the action of the  structure group $G$, where $V_{u}$ is the \emph{vertical subspace} of $T_{u}\left(P\right)$, namely the tangent subspace of the fiber $G_{p}$ in $u$. Given a connection $\Gamma$ in $P$, we define a 1-form $\omega\in  \mathfrak{g}\otimes T^{*}P$ with values in the Lie algebra $\mathfrak{g}$ of Lie group $G$ called \emph{connection form}.  In other words, for each  $X\in T_{u}\left(P\right)$, one can  define $\omega\left(X\right)$ to be the unique $A\in \mathfrak{g}$ such that $\left(A^{*}\right)_{u}$, the  fundamental vector field generated by $A$, is equal to the vertical component of $X$ (see also \cite{Ali}).  
In this perspective, the connection is a rule that  allows to compare vector fields belonging to different vector spaces  on the manifold \cite{KN,LEE, LEER, MI, AT, NAK, SGMMP, YCB}.  

In TEGR,  dynamical  variables   are not the metric components    $\bf g$, as in GR, but tetrads.  These  tetrad fields, the {\it  vierbeins}  $h_{a}\left(x^{\mu}\right)$, are orthonormal vector fields on   $\cal M$ defining a basis   for any point $p\in \cal M$  with coordinates $x^{\mu}$.  The  tangent space  to $\cal M$ is  Minkowski with  metric  $\eta_{ab}=diag\left(1,-1,-1,-1\right)$.

One can express the  tetrad basis \{$h_{a}\}$ and its  dual \{$h^{a}\}$ in terms of the holonomic coordinate basis  $\{e_{\mu}\}=\{\partial_{\mu}\}$ and its   dual basis  $\{e^{\mu}\}=\{dx^{\mu}\}$  \cite{CR, WGR, GRG, HS}. We have
\begin{equation}
h_{a}=h_{a}^{\phantom{a}\mu}e_{\mu}h^{a}=h^{a}_{\phantom{a}\mu}e^{\mu}\,,
\end{equation}
\begin{equation}
\eta_{ab}=g_{\mu\nu}h_{a}^{\phantom{a}\mu}h_{b}^{\phantom{a}\nu}  g_{\mu\nu}=\eta_{ab}h^{a}_{\phantom{a}{\mu}}h_{b}^{\phantom{b}\nu}\,,
\end{equation}
\begin{equation}
h^{a}_{\phantom{a}{\mu}}h_{a}^{\phantom{a}{\nu}}=\delta_{\mu}^{\nu}h^{a}_{\phantom{a}\mu}h_{b}^{\phantom{b}\mu}=\delta_{a}^{b}\,.
\end{equation}
A \emph{connection form} $\omega$ (\emph{spin connection} or \emph{Lorentz connection}) of the given connection $\Gamma$ in the bundle of linear frames, i.e. a \emph{linear connection}, is a 1-form $\omega$ with values in the Lie algebra of Lorentz group $\mathfrak{so}\left(1,3\right)$:
\begin{equation}
\omega=\omega_{\mu}dx^{\mu}=\frac{1}{2}\omega^{a}_{\phantom{a}b}S_{a}^{\phantom{a}b}=\frac{1}{2}\omega^{a}_{\phantom{a}b\mu}S_{a}^{\phantom{a}b}dx^{\mu}\,,
\end{equation}
with $S_{a}^{\phantom{a}b}$ an appropriate representation of the Lorentz generators and $\omega^{a}_{\phantom{a}b}\in A^{1}\left(\cal M\right)$, where $\omega^{a}_{\phantom{a}b}$ are matrices of 1-forms and $A^{1}\left(\cal M\right)$,  the space of all $1$-forms \cite{APTG, AP, SN}. 
Defining the  \emph{absolute exterior differential} or the \emph{covariant exterior derivative} of a given  tensor $\left(r,s\right)$, valued on the $p$-forms $B^{a}_{\phantom{a}b}\in A^{p}\left({\cal M},\mathcal{T}^{r}_{s}\left({\cal M}\right)\right)$, as the operator $D:A^{p}\left({\cal M},\mathcal{T}^{r}_{s}\right)\rightarrow A^{p+1}\left({\cal M},\mathcal{T}^{r}_{s}\right)$, we have
\begin{equation}
DB^{a}_{\phantom{a}b}=dB^{a}_{\phantom{a}b}+\omega^{a}_{\phantom{a}c}\wedge B^{c}_{\phantom{c}b}-\omega^{d}_{\phantom{d}b}\wedge B^{a}_{\phantom{a}d}\,,
\end{equation}
with the operator $d:A^{p}\left({\cal M}\right)\rightarrow A^{p+1}\left({\cal M}\right)$ the \emph{exterior derivative} of a $p$-form. Immediately, we obtain the \emph{Cartan structure equations}:
\begin{gather}
T^{a}=Dh^{a}=dh^{a}+\omega^{a}_{\phantom{a}b}\wedge h^{b}=\frac{1}{2}T^{a}_{\phantom{a}bc}h^{b}\wedge h^{c}\,,\\
R^{a}_{\phantom{a}b}=D\omega^{a}_{\phantom{a}b}=d\omega^{a}_{\phantom{a}b}+\omega^{a}_{\phantom{a}c}\wedge\omega^{c}_{\phantom{c}b}=\frac{1}{2}R^{a}_{\phantom{a}bcd}h^{c}\wedge h^{d}\,,
\end{gather}
where $T^{a}$ and $R^{a}_{\phantom{a}b}$ are the \emph{torsion} and \emph{curvature 2-forms}. The T and R forms are the Lie algebras of the 2-forms valued on the translations  and Lorentz groups respectively 
\begin{equation}
\textnormal{T}=T^{a}P_{a}=\frac{1}{2}T^{a}_{\phantom{a}\mu\nu}P_{a}dx^{\mu}\wedge dx^{\nu}\,,
\end{equation} 
\begin{equation}
\textnormal{R}=\frac{1}{2}R^{a}_{\phantom{a}b}S_{a}^{\phantom{a}b}=\frac{1}{4}R^{a}_{\phantom{a}b\mu\nu}S_{a}^{\phantom{a}b}dx^{\mu}\wedge dx^{\nu}\,,
\end{equation}
relative to a coordinate basis $\{dx^{\mu}\}$ where $P_{a}$ are the translation generators and, as said above, 
$S_{a}^{\phantom{a}b}$ are the Lorentz generators. Furthermore, torsion and curvature forms satisfy the following Bianchi identities:
\begin{align}
DT^{a}&=R^{a}_{\phantom{a}b}\wedge h^{b}&\text{First Bianchi Identity}\,,\\
DR^{a}_{\phantom{a}b}&=0&\text{Second Bianchi Identity}\,.
\end{align}
Equivalently,  let $\left\{\cal M,\pi,\cal E\right\}$ be a vector bundle on a manifold $\cal M$.  We can define a connection as a bilinear map $\nabla : \mathcal{T}\left(\cal M\right)\times \mathcal{E}\left(\cal M\right)\rightarrow\mathcal{E}\left(\cal M\right)$, namely $\left(X,V\right)\mapsto \nabla_{X}V$, such that is \emph{$C^{\infty}\left(\cal M\right)$-linear} in $X\in \mathcal{T}\left(\cal M\right)$, \emph{$\mathbb{R}$-linear} in $V\in \mathcal{E}\left(\cal M\right)$ and the Leibniz formula is satisfied  for $\mathcal{T}\left(\cal M\right)$,  the set of all vector fields, and $\mathcal{E}\left(\cal M\right)$, the space of smooth sections of the vector bundle. The \emph{covariant derivative} $\nabla_{X}V$ of $V$ along $X$ is the section of map $\nabla$. Thus the \emph{connection coefficients}  $\omega^{a}_{\phantom{a}bc}$ of connection $\omega$ expressed in an arbitrary anholonomic basis of tetrads $\{h_{a}\}$:
\begin{equation}
dh^{a}=-\frac{1}{2}f^{a}_{\phantom{a}bc}h^{b}\wedge h^{c}\Longleftrightarrow \left[h_{b},h_{c}\right]=f^{a}_{\phantom{a}bc}h_{a}\,,
\end{equation}
\begin{equation}
\nabla_{h_{a}}h_{b}=\omega^{c}_{\phantom{c}a}\left(h_{b}\right)h_{c}=\omega^{c}_{\phantom{c}ba}h_{c}\,.
\end{equation}
From the formula of 1-form \emph{exterior derivative}  
\begin{equation}
d\omega\left(X,Y\right)=X\omega\left(Y\right)-Y\omega\left(X\right)-\omega\left(\left[X,Y\right]\right)\,,
\end{equation}
we get
\begin{gather}
\tau\left(X,Y\right)=T^{a}\left(X,Y\right)h_{a}\,,\\
\mathcal{R}\left(X,Y\right)h_{b}=R^{a}_{\phantom{a}b}\left(X,Y\right)h_{a}\,,
\end{gather}
which define the torsion and curvature of  linear connection  $\omega$ on $\cal M$. A map $\tau : \mathcal{T}\left(\cal M\right)\times\mathcal{T}(\cal M)\rightarrow \mathcal{T}(\cal M)$ :
\begin{equation}
\tau\left(X,Y\right)=\nabla_{X}Y-\nabla_{Y}X-\left[X,Y\right]\,,
\end{equation}
with $\tau$ a tensor field of type $\left(1,2\right)$ is  the {\it torsion tensor}. The {\it Riemann (curvature) tensor} R is a map $\textnormal{R} : \mathcal{T}\left(\cal M\right)\times\mathcal{T}(\cal M)\times\mathcal{T}\left(\cal M\right)\rightarrow \mathcal{T}(\cal M)$ defined as:
\begin{equation}
\textnormal{R}\left(X,Y,Z\right)=\mathcal{R}\left(X,Y\right)Z=\nabla_{X}\nabla_{Y}Z-\nabla_{Y}\nabla_{X}Z-\nabla_{\left[X,Y\right]}Z\,,
\end{equation}
with R a  tensor field of type $\left(1,3\right)$. 
The connection coefficients of $\omega$ on $\cal M$, in terms of the only coordinate bases, are traditionally referred to as \emph{Christoffel symbols} $\Gamma^{\alpha}_{\phantom{\alpha}\mu\nu}$, that are related to the connection coefficients $\omega^{a}_{\phantom{a}b\mu}$ in mixed bases by:
\begin{equation}
\nabla_{e_{\mu}}e_{\nu}=\Gamma^{\alpha}_{\phantom{\alpha}\nu\mu}e_{\alpha}\,,\qquad \nabla_{e_{\mu}}h_{b}=\omega^{a}_{\phantom{a}b\mu}h_{a}\,,
\end{equation}
\begin{equation}\label{Gammaversusomega}
\Gamma^{\alpha}_{\phantom{\alpha}\nu\mu}=h_{a}^{\phantom{a}\alpha}\partial_{\mu}h^{a}_{\phantom{a}\nu}+h_{a}^{\phantom{a}\alpha}\omega^{a}_{\phantom{a}b\mu}h^{b}_{\phantom{b}\nu}\,.
\end{equation}
Eq. \eqref{Gammaversusomega} is nothing else but  the  \emph{tetrad postulate} or the \emph{absolute parallelism}, that  is:
\begin{equation}
\nabla_{\mu}h^{a}_{\phantom{a}\rho}=\partial_{\mu}h^{a}_{\phantom{a}\rho}-\Gamma^{\alpha}_{\phantom{\alpha}\rho\mu}h^{a}_{\phantom{a}\alpha}+\omega^{a}_{\phantom{a}b\mu}h^{b}_{\phantom{b}\rho}=0\,.
\end{equation}
It states, essentially, that  the vierbiens are parallel vector fields. We can derive the components of torsion tensor $\tau$ and curvature tensor $\textnormal{R}$  of an arbitrary connection $\omega$ with respect to anholomic frames of the tetrad fields $\{h_{a}\}$ as:
\begin{equation}
\tau\left(h_{a},h_{b}\right)=T^{c}_{\phantom{c}ab}h_{c}=\left(\omega^{c}_{\phantom{c}ba}-\omega^{c}_{\phantom{c}ab}-f^{c}_{\phantom{c}ab}\right)h_{c}\,,
\end{equation}
\begin{equation}
\mathcal{R}\left(h_{a},h_{b}\right)h_{c}=R^{d}_{\phantom{d}cab}h_{d}=\left(h_{a}\left(\omega^{d}_{\phantom{d}cb}\right)-h_{b}\left(\omega^{d}_{\phantom{d}ca}\right)+\omega^{e}_{\phantom{e}cb}\omega^{d}_{\phantom{d}ea}-\omega^{e}_{\phantom{e}ca}\omega^{d}_{\phantom{d}eb}-f^{g}_{\phantom{g}ab}\omega^{d}_{\phantom{d}cg}\right)h_{d}\,.
\end{equation} 
The torsion and curvature components in the mixed algebraic and spacetime indices are:
 \begin{equation}
T^{a}_{\phantom{a}\nu\mu}=\partial_{\nu}h^{a}_{\phantom{a}\mu}-\partial_{\mu}h^{a}_{\phantom{a}\nu}+\omega^{a}_{\phantom{a}e\nu}h^{e}_{\phantom{e}\mu}-\omega^{a}_{\phantom{a}e\mu}h^{e}_{\phantom{e}\nu}\,,
\end{equation}
\begin{equation}
R^{a}_{\phantom{a}b\nu\mu}=\partial_{\nu}\omega^{a}_{\phantom{a}b\mu}-\partial_{\mu}\omega^{a}_{\phantom{a}b\nu}+\omega^{a}_{\phantom{a}e\nu}\omega^{e}_{\phantom{e}b\mu}-\omega^{a}_{\phantom{a}e\mu}\omega^{e}_{\phantom{e}b\nu}\,.
\end{equation}
On the other hand, the torsion and curvature components, in a natural basis, are given respectively by 
\begin{equation}
\tau\left(e_{\mu},e_{\nu}\right)=T^{\rho}_{\phantom{\rho}\mu\nu}e_{\rho}=\left(\Gamma^{\rho}_{\phantom{\rho}\nu\mu}-\Gamma^{\rho}_{\phantom{\rho}\mu\nu}\right)e_{\rho}\,,
\end{equation}
\begin{equation}
\mathcal{R}\left(e_{\lambda},e_{\nu}\right)e_{\mu}=R^{\rho}_{\phantom{\rho}\mu\lambda\nu}e_{\rho}=\left(\partial_{\lambda}\Gamma^{\rho}_{\phantom{\rho}\mu\nu}-\partial_{\nu}\Gamma^{\rho}_{\phantom{\rho}\mu\lambda}+\Gamma^{\rho}_{\phantom{\rho}\sigma\lambda}\Gamma^{\sigma}_{\phantom{\sigma}\mu\nu}-\Gamma^{\rho}_{\phantom{\rho}\sigma\nu}\Gamma^{\sigma}_{\phantom{\sigma}\mu\lambda}\right)e_{\rho}\,.
\end{equation}
The first Bianchi identity for torsion, in spacetime components,  is:
\begin{equation}
\begin{split}
\nabla_{\nu}T^{\lambda}_{\phantom{\lambda}\rho\mu}+\nabla_{\mu}T^{\lambda}_{\phantom{\lambda}\nu\rho}+\nabla_{\rho}T^{\lambda}_{\phantom{\lambda}\mu\nu}&\\
=R^{\lambda}_{\phantom{\lambda}\rho\mu\nu}+R^{\lambda}_{\phantom{\lambda}\nu\rho\mu}+R^{\lambda}_{\phantom{\lambda}\mu\nu\rho}+&T^{\lambda}_{\phantom{\lambda}\rho\sigma}T^{\sigma}_{\phantom{\sigma}\mu\nu}+T^{\lambda}_{\phantom{\lambda}\nu\sigma}T^{\sigma}_{\phantom{\sigma}\rho\mu}+T^{\lambda}_{\phantom{\lambda}\mu\sigma}T^{\sigma}_{\phantom{\sigma}\nu\rho}\,.
\end{split}
\end{equation}
The second Bianchi identity for curvature in spacetime components is:
\begin{equation}
\begin{split}
\nabla_{\nu}R^{\lambda}_{\phantom{\lambda}\sigma\rho\mu}+\nabla_{\mu}R^{\lambda}_{\phantom{\lambda}\sigma\nu\rho}+\nabla_{\rho}R^{\lambda}_{\phantom{\lambda}\sigma\mu\nu}\\
=R^{\lambda}_{\phantom{\lambda}\sigma\mu\theta}T^{\theta}_{\phantom{\theta}\nu\rho}+R^{\lambda}_{\phantom{\lambda}\sigma\nu\theta}T^{\theta}_{\phantom{\theta}\rho\mu}+R^{\lambda}_{\phantom{\lambda}\sigma\rho\theta}T^{\theta}_{\phantom{\theta}\mu\nu}\,,
\end{split}
\end{equation}
where $\nabla_{\mu}$ is the   covariant derivative along the   coordinate basis $e_{\mu}$ with respect to the   generic connection $\omega$. In presence of torsion, the  covariant derivative commutator of a scalar function   does not   commute as in GR. We have, for a scalar function and a vector field, the relations:
\begin{equation}
\left[\nabla_{\mu},\nabla_{\nu}\right]f=-T^{\rho}_{\phantom{\rho}\mu\nu}\nabla_{\rho}f\,,
\end{equation}
\begin{equation}
\left[\nabla_{\mu},\nabla_{\nu}\right]V^{\rho}=R^{\rho}_{\phantom{\rho}\alpha\mu\nu}V^{\alpha}-T^{\sigma}_{\phantom{\sigma}\mu\nu}\nabla_{\sigma}V^{\rho}\,.
\end{equation}
The  Weitzenb\"ock connection is curvature-free and  metric compatible. In a coordinate frame, it is defined in terms of a particular tetrad where the Lorentz connection $\omega^{a}_{\phantom{a}b\mu}$ is vanishing. Then we have: 
\begin{equation}
\tilde{\Gamma}^{\rho}_{\phantom{\rho}\mu\nu}\equiv h_{a}^{\phantom{a}\rho}\partial_{\nu}h^{a}_{\phantom{a}\mu}=-h^{a}_{\phantom{a}\mu}\partial_{\nu}h_{a}^{\phantom{a}\rho}\,,
\end{equation}
satisfying the  metricity condition
\begin{equation}
\tilde{\nabla}_{\lambda}g_{\mu\nu}\equiv \partial_{\lambda}g_{\mu\nu}-\tilde{\Gamma}^{\rho}_{\phantom{\rho}\lambda\mu}g_{\rho\nu}-\tilde{\Gamma}^{\rho}_{\phantom{\rho}\lambda\nu}g_{\mu\rho}=0\,,
\end{equation}
and the absolute parallelism postulate or  tetrad postulate
\begin{equation}
\tilde{\nabla}_{\lambda}h^{a}_{\phantom{a}\mu}=\partial_{\lambda}h^{a}_{\phantom{a}\mu}-\tilde{\Gamma}^{\rho}_{\phantom{\rho}\mu\lambda}h^{a}_{\phantom{a}\rho}=0\,.
\end{equation}
Torsion and vanishing curvature components of Weitzenb\"ock connection in a coordinate basis are:
\begin{equation}
T^{\rho}_{\phantom{\rho}\mu\nu}\equiv \tilde{\Gamma}^{\rho}_{\phantom{\rho}\nu\mu}-\tilde{\Gamma}^{\rho}_{\phantom{\rho}\mu\nu}=h_{a}^{\phantom{a}\rho}\partial_{\mu}h^{a}_{\phantom{a}\nu}-h_{a}^{\phantom{a}\rho}\partial_{\nu}h^{a}_{\phantom{a}\mu}\quad \tilde{R}^{\rho}_{\mu\lambda\nu}=0\,,
\end{equation} 
\begin{equation}
T^{\rho}_{\phantom{\rho}\left(\mu\nu\right)}=0\,.
\end{equation}
 $\tilde{\nabla}$ is the covariant derivative relative to the Weitzenb\"ock connection. The Weitzenb\"ock connection satisfies both  the metricity condition and the  tetrad postulate. In terms of  parallelism,  choosing the  Weitzenb\"ock connection has a  straightforward interpretation: If we perform the covariant  derivative of a generic   vector field  $V=V^{\mu}e_{\mu}=V^{a}h_{a}$ with respect to a vector field  $X=X^{\nu}e_{\nu}$, we have: 
\begin{gather*}
\tilde{\nabla}_{X}V=X^{\nu}\left[\partial_{\nu}V^{\alpha}+V^{\mu}\tilde{\Gamma}^{\alpha}_{\phantom{\alpha}\mu\nu}\right]e_{\alpha}=X^{\nu}\left[\partial_{\nu}V^{\alpha}+h_{a}^{\phantom{a}\alpha}\partial_{\nu}h^{a}_{\phantom{a}\mu}V^{\mu}\right]e_{\alpha}=X^{\nu}\left[h_{a}^{\phantom{a}\alpha}\partial_{\nu}V^{a}\right]e_{\alpha}\\
\tilde{\nabla}_{X}V=0 \quad \Rightarrow \quad \partial_{\nu}V^{a}=0\,,
\end{gather*}
that is the vector $V$ is parallel transported by the  Weitzenb\"ock connection along the vector field  $X$, if its components along the tetrad basis are constant.  This means that the  tetrad field  "parallelizes" the  spacetime. The  Latin indices  $a,b,c\dots$ are the "flat"   or  holonomic indices. They refer to  tensor objects  projected by  tetrads on the  tangent space. They can be raised or lowered by the  Minkowski tensor   $\eta_{ab}$, that is  $F_{a}=\eta_{ab}F^{b}$. The Greek  indices   $\mu,\nu,\dots$, called the  "curved"  or anholonomic indices,  refer to tensor objects defined on the  Riemannian manifod. They can be raised and lowered  by the  metric $g_{\mu\nu}$. In summary, the  tetrad $h^{a}_{\phantom{a}\rho}$ and its dual  $h_{a}^{\phantom{a}\rho}$ allows to  project a geometric object   from the  Riemann manifold  to the tangent space and  viceversa, that is  $F^{a}=h^{a}_{\phantom{a}\rho}F^{\rho}$ and  $F^{\rho}=h_{a}^{\phantom{a}\rho}F^{a}$. 
The  difference between the  torsionless Levi-Civita connection and  the  Weitzenb\"ock  one is given by the  \emph{contortion tensor}:
\begin{equation}
\label{contor}
K^{\rho}_{\phantom{\rho}\mu\nu}\equiv\tilde{\Gamma}^{\rho}_{\phantom{\rho}\mu\nu}-\stackrel{\circ}\Gamma{}^{\rho}_{\phantom{\rho}\mu\nu}=-\frac{1}{2}\left(T^{\rho}_{\phantom{\rho}\mu\nu}-T_{\nu\phantom{\rho}\mu}^{\phantom{\mu}\rho}+T_{\mu\nu}^{\phantom{\mu\nu}\rho}\right)=h_{a}^{\phantom{a}\rho}\nabla_{\nu}h^{a}_{\phantom{a}\mu}\,,
\end{equation}
\begin{equation}
K^{\left(\rho\mu\right)\nu}=0\,,
\end{equation}
where 
\begin{equation}
\stackrel{\circ}\Gamma{}^{\rho}_{\phantom{\rho}\mu\nu}=\frac{1}{2}g^{\sigma\rho}\left(g_{\rho\mu,\nu}+g_{\rho\nu,\mu}-g_{\mu\nu,\rho}\right)\,,
\end{equation}
is the  Levi Civita torsion-free connection and  $\nabla_{\mu}$ is the   covariant derivative with  respect to the  Levi Civita connection.
We can define the  \emph{superpotential} tensor as:
\begin{equation}
S^{\rho\mu\nu}\equiv \frac{1}{2}\left(K^{\mu\nu\rho}-g^{\rho\nu}T^{\sigma\mu}_{\phantom{\sigma\mu}\sigma}+g^{\rho\mu}T^{\sigma\nu}_{\phantom{\sigma\nu}\sigma}\right)\,,
\end{equation}
\begin{equation}
S^{\rho\left(\mu\nu\right)}=0\,,
\end{equation}
\begin{equation}
K^{\mu}_{\phantom{\mu}\rho\mu}=-T^{\mu}_{\phantom{\mu}\rho\mu}=S^{\mu}_{\phantom{\mu}\rho\mu}\,.
\end{equation}
The  scalar torsion $T$ is then:
\begin{equation}
\begin{split}
T&\equiv S^{\rho\mu\nu}T_{\rho\mu\nu}=-2S^{\rho\mu\nu}K_{\rho\mu\nu}\,,\\
&=\frac{1}{4}T^{\rho\mu\nu}T_{\rho\mu\nu}+\frac{1}{2}T^{\rho\mu\nu}T_{\nu\mu\rho}-T^{\rho}_{\phantom{\rho}\mu\rho}T^{\nu\mu}_{\phantom{\nu\mu}\nu}\,,\\
&=K^{\mu\nu\rho}K_{\rho\nu\mu}-K^{\mu\rho}_{\phantom{\mu\rho}\mu}K^{\nu}_{\phantom{\nu}\rho\nu}\,.
\end{split}
\end{equation}
The Riemann tensor for the  Weitzenb\"ock connection is null:
\begin{equation}
\textnormal{R}\left[\tilde{\Gamma}\right]=0\,,
\end{equation}
and then we obtain  the  relation between  the spacetime   components of    curvature tensor in the Weitzenb\"ock and  Levi-Civita representations, that is:
\begin{equation}
0=\tilde{R}^{\rho}_{\phantom{\rho}\mu\lambda\nu}=-\stackrel{\circ}{R^{\rho}}_{\mu\lambda\nu}+\nabla_{\nu}K^{\rho}_{\phantom{\rho}\mu\lambda}-\nabla_{\lambda}K^{\rho}_{\phantom{\rho}\mu\nu}+K^{\rho}_{\phantom{\rho}\sigma\nu}K^{\sigma}_{\phantom{\sigma}\mu\lambda}-K^{\rho}_{\phantom{\rho}\sigma\lambda}K^{\sigma}_{\phantom{\sigma}\mu\nu}\,.
\end{equation}
The Ricci tensor is obtained by $R_{\mu\nu}=R^{\rho}_{\phantom{\rho}\mu\rho\nu}$ and then:
\begin{equation}\label{TENSRICC}
\begin{split}
R_{\mu\nu}&=\nabla_{\nu}K^{\rho}_{\phantom{\rho}\mu\rho}-\nabla_{\rho}K^{\rho}_{\phantom{\rho}\mu\nu}+K^{\rho}_{\phantom{\rho}\sigma\nu}K^{\sigma}_{\phantom{\sigma}\mu\rho}-K^{\rho}_{\phantom{\rho}\sigma\rho}K^{\sigma}_{\phantom{\sigma}\mu\nu}\\
&=-2\nabla^{\rho}S_{\nu\rho\mu}-g_{\mu\nu}\nabla^{\rho}T^{\sigma}_{\phantom{\sigma}\rho\sigma}-2S^{\rho\sigma}_{\phantom{\rho\sigma}\mu}K_{\sigma\rho\nu}\,.
\end{split}
\end{equation}
Contracting again, we get  the Ricci scalar expressed with respect to the  Levi Civita connection
\begin{equation}\label{Scalare di curvat}
R\left[\stackrel{\circ}\Gamma\right]\left(h\right)=-T-\frac{2}{h}\partial_{\mu}\left(hT^{\nu\mu}_{\phantom{\nu\mu}\nu}\right)=-T-2\nabla_{\mu}T^{\mu}\,,
\end{equation}
where  $h=det\left(h^{a}_{\phantom{a}\mu}\right)=\sqrt{-g}$ and $T^{\mu}=T^{\nu\mu}_{\phantom{\nu\mu}\nu}$ is the  contraction of torsion tensor with respect to the first and the third index\footnote{In Eq.\eqref{Scalare di curvat}, the boundary term is defined with the minus sign according to the definition of contortion tensor \eqref{contor}. In Ref.\cite{BC}, the relation between the curvature and torsion scalar is $R=-T+B$ due to a different definition of the contortion tensor. However, the two approaches are equivalent.}.  
In  TEGR,  the dynamical variables are  the  16 vierbien components  $h^{a}_{\phantom{a}\mu}$ instead of the  10 metric components  $g_{\alpha\beta}$ of GR. If we choose a tetrad basis  to annul the Lorentz connection $\omega^{a}_{\phantom{a}b}=0$, we lose  the local Lorentz invariance and preserve the global one.   Then we have  16 equations instead of   10 as in GR.  If we restore the local Lorentz invariance by a non-trivial spin connection, it means  that  just 10 of the  total 16 tetrad components are  independents while the other 6 are fixed by the gauge.  In absence of such a local symmetry, we cannot choose  the  tetrad  unless than a global Lorentz transformation. 

\section{The gravitational energy-momentum pseudotensor of $f\left(R\right)$ gravity}\label{GEMTR}
Let us consider the action:
\begin{equation}
\mathcal{S}_{f\left(R\right)}=\frac{1}{2\kappa^{2}}\int_{\Omega}d^{4}x\sqrt{-g}f\left(R\right)\,,
\end{equation}
with $\kappa^{2}=8\pi G/c^{4}$. We can calculate the  variation $\tilde{\delta}$ with respect to the metric  $g^{\mu\nu}$ and  coordinates $x^{\mu}$ for a generic infinitesimal  transformation:
\begin{equation}
x^{\prime\mu}=x^{\mu}+\delta x^{\mu}\,,\qquad g^{\prime\mu\nu}\left(x^{\prime}\right)= g^{\mu\nu}\left(x\right)+\tilde{\delta}g^{\mu\nu}\,,\qquad g^{\prime\mu\nu}\left(x\right)= g^{\mu\nu}\left(x\right)+\delta g^{\mu\nu}\,,
\end{equation}
\begin{equation}\label{varloc}
\tilde{\delta}\mathcal{S}_{f\left(R\right)}=\frac{1}{2\kappa^{2}}\int_{\Omega}d^{4}x\left[\delta\left(\sqrt{-g}f\left(R\right)\right)+\partial_{\mu}\left(\sqrt{-g}f\left(R\right) \delta x^{\mu}\right)\right]\,,
\end{equation}
where $\tilde{\delta}$ is the local  variation and   $\delta$ is the global  variation for a fixed $x$.
We obtain \cite{CL, CF, HE, SGMM, OR, BR}:
\begin{equation}\label{varlocfR}
\begin{split}
\tilde{\delta}\mathcal{S}_{f\left(R\right)}=\frac{1}{2\kappa^{2}}\int_{\Omega}d^{4}x\sqrt{-g}\left[f'\left(R\right)R_{\mu\nu}-\frac{1}{2}g_{\mu\nu}f\left(R\right)-\nabla_{\mu}\nabla_{\nu}f'\left(R\right)+g_{\mu\nu}\Box f'\left(R\right)\right]\delta g^{\mu\nu}\\
+\int_{\Omega}d^{4}x\partial_{\alpha}\Biggl\{\frac{\sqrt{-g}}{2\kappa^{2}}\biggl[\partial_{\beta}f'\left(R\right)\left(g^{\eta\rho}g^{\alpha\beta}-g^{\alpha\eta}g^{\rho\beta}\right)\delta g_{\eta\rho}+f'\left(R\right)\Bigl[\left(\stackrel{\circ}\Gamma{}^{\rho\eta\alpha}-\stackrel{\circ}\Gamma{}^{\eta\sigma}{}_{\sigma}g^{\alpha\rho}\right)\delta g_{\eta\rho}\\
+\left(g^{\alpha\eta}g^{\tau\rho}-g^{\eta\rho}g^{\alpha\tau}\right)\delta g_{\eta\rho,\tau}\Bigr]+f\left(R\right)\delta_{\lambda}^{\alpha}\delta x^{\lambda}\biggr]\Biggr\}\,,
\end{split}
\end{equation}
where $f'\left(R\right)=\partial f/\partial R$. Imposing the action stationarity   at a fixed  $x$, that is  $\delta\mathcal{S}_{f\left(R\right)}=0$ in a given domain $\Omega$ where the total variation of  both  metric and its first derivatives are zero at the boundary, that is   $\delta g_{\mu\nu}\vert_{\partial\Omega}=0$ and  $\delta \left(\partial_{\alpha}g_{\mu\nu}\right)\vert_{\partial\Omega}=0$,  the field equations in vacuum are:
\begin{equation}\label{ECFRV}
-P^{f\left(R\right)}_{\mu\nu}=\frac{2\kappa^{2}}{\sqrt{-g}}\frac{\delta L_{f\left(R\right)}}{\delta g^{\mu\nu}}=f'\left(R\right)R_{\mu\nu}-\frac{1}{2}g_{\mu\nu}f\left(R\right)-\nabla_{\mu}\nabla_{\nu}f'\left(R\right)+g_{\mu\nu}\Box f'\left(R\right)=0\,,
\end{equation}
where  $2\kappa^{2}L_{f\left(R\right)}=\sqrt{-g}f\left(R\right)$. For an infinitesimal transformation like a rigid translation, one has:
\begin{equation}\label{TrRig}
x^{\prime\mu}=x^{\mu}+\epsilon^{\mu}\Rightarrow \delta g_{\mu\nu}=-\epsilon^{\lambda}g_{\mu\nu,\lambda}\,, 
\end{equation}
because $\partial \epsilon=0$.  If the local variation of the action is zero and the field   $g_{\mu\nu}$ satisfies  the field equations, we have: 
\begin{equation}
\tilde{\delta}\mathcal{S}_{f\left(R\right)}=0 \Rightarrow \partial_{\sigma}\left(\sqrt{-g}\tau^{\sigma}_{\phantom{\sigma}{\lambda | f\left(R\right)}}\right)=0\,,
\end{equation}
with
\begin{equation}
\boxed{
\begin{split}
2\kappa^{2}\tau^{\sigma}{}_{\lambda | f\left(R\right)}=&2\partial_{\beta}f'\left(R\right)g^{\eta[\rho}g^{\sigma]\beta}g_{\eta\rho,\lambda}\\&+f'\left(R\right)\Bigl[\bigl(\stackrel{\circ}{\Gamma}{}^{\rho\eta\sigma}
-\stackrel{\circ}{\Gamma}{}^{\eta\alpha}_{\phantom{\eta\alpha}{\alpha}}g^{\sigma\rho}\bigr)g_{\eta\rho,\lambda}+2g^{\sigma[\eta}g^{\tau]\rho}g_{\eta\rho,\tau\lambda}\Bigr]-f\left(R\right)\delta_{\lambda}^{\sigma}
\end{split}
}
\end{equation}
This is the \emph{gravitational energy-momentum pseudotensor of $f\left(R\right)$ gravity},   with \\$\stackrel{\circ}{\Gamma}{}^{\rho\eta\sigma}=g^{\eta\epsilon}g^{\sigma\varphi}\stackrel{\circ}{\Gamma}{}^{\rho}_{\phantom{\eta}{\epsilon\varphi}}$, and  $\stackrel{\circ}{\Gamma}{}^{\eta\alpha}_{\phantom{\eta\alpha}{\alpha}}=g^{\alpha\epsilon}\stackrel{\circ}{\Gamma}{}^{\eta}_{\phantom{\eta}{\epsilon\alpha}}$.
Considering also matter fields, we have:
\begin{equation}
\mathcal{S}_{m}=\int_{\Omega}d^{4}x L_{m}\,,
\end{equation}
where  $L_{m}$ depends, at most,  on first  derivatives of metric $g_{\mu\nu}$. Imposing the same variational conditions as above, we have:
\begin{equation}
\delta\mathcal{S}_{m}=\int_{\Omega}d^{4}x \frac{\delta L_{m}}{\delta g^{\mu\nu}}\delta g^{\mu\nu}=\int_{\Omega}d^{4}x \left(\frac{\sqrt{-g}}{2}\right)T^{\left(m\right)}_{\mu\nu}\delta g^{\mu\nu}\,,
\end{equation}
where $T^{\left(m\right)}_{\mu\nu}$ is the energy-momentum tensor of matter fields:
\begin{equation}
T^{\left(m\right)}_{\mu\nu}=\frac{2}{\sqrt{-g}}\frac{\delta L_{m}}{\delta g^{\mu\nu}}\,.
\end{equation}
In summary, by  minimizing the total action  $\mathcal{S_{T}}=\mathcal{S}_{f\left(R\right)}+\mathcal{S}_{m}$, one  obtains the following  field equations in  presence of matter:
\begin{equation}\label{ECFRM}
{}^{f\left(R\right)}P_{\mu\nu}=\kappa^{2}T^{\left(m\right)}_{\mu\nu}\,.
\end{equation}
From the contracted Bianchi identities, we have:
\begin{equation}\label{IBFR}
\nabla^{\nu}G_{\mu\nu}=0\leftrightarrow\nabla^{\nu} \left({}^{f\left(R\right)}P_{\mu\nu}\right)=0\leftrightarrow\nabla^{\nu}T^{\left(m\right)}_{\mu\nu}=0\,,
\end{equation}
where we adopted the formula  \eqref{trinabla} in Appendix $A$. Finally, from the variation \eqref{varlocfR} for a rigid translation  \eqref{TrRig} and from the matter field equations  \eqref{ECFRM}, we have:
\begin{equation}
\begin{split}
\delta L_{f\left(R\right)}+\partial_{\sigma}\left(L_{f\left(R\right)}\delta x^{\sigma}\right)&=\frac{\sqrt{-g}}{2\kappa^{2}}P_{{f\left(R\right)}}^{\mu\nu}\delta g_{\mu\nu}-\partial_{\sigma}\left(\sqrt{-g}\tau^{\sigma}_{\phantom{\sigma}{\lambda}}\right)\epsilon^{\lambda}\\&=\left[-\frac{1}{2}\sqrt{-g}T_{\left(m\right)}^{\mu\nu}g_{\mu\nu,\lambda}-\partial_{\sigma}\left(\sqrt{-g}\tau^{\sigma}_{\phantom{\sigma}{\lambda}}\right)\right]\epsilon^{\lambda}\,.
\end{split}
\end{equation}
From the following identity which holds because  $T^{\eta}_{\phantom{\eta}\alpha}$ is  symmetric, one has:
\begin{equation}
\sqrt{-g}\nabla_{\eta}T^{\eta}_{\phantom{\eta}{\alpha}}=\partial_{\eta}\left(\sqrt{-g}T^{\eta}_{\phantom{\sigma}{\alpha}}\right)-\frac{1}{2}g_{\rho\sigma,\alpha}T^{\rho\sigma}\sqrt{-g}\,.
\end{equation}  
It is:
\begin{equation}
\delta L_{f\left(R\right)}+\partial_{\sigma}\left(L_{f\left(R\right)}\delta x^{\sigma}\right)=\left[-\partial_{\sigma}\left(\sqrt{-g}T^{\sigma}_{\phantom{\sigma}{\lambda}}\right)+\sqrt{-g}T^{\sigma}_{\phantom{\sigma}{\lambda;\sigma}}-\partial_{\sigma}\left(\sqrt{-g}\tau^{\sigma}_{\phantom{\sigma}{\lambda}}\right)\right]\epsilon^{\lambda}\,.
\end{equation}
If, for rigid translations, the local variation is zero, one has:
\begin{equation}
\delta L_{f\left(R\right)}+\partial_{\sigma}\left(L_{f\left(R\right)}\delta x^{\sigma}\right)=0 \rightarrow \partial_{\sigma}\left[\sqrt{-g}\left(\tau^{\sigma}_{\ \lambda}+T^{\sigma}{}_{\lambda}\right)\right]=\sqrt{-g}\nabla_{\sigma}T^{\sigma}{}_{\lambda}\,,
\end{equation}
and then, from the contracted Bianchi identity  \eqref{IBFR}, we obtain the total conservation law:
\begin{equation}
\partial_{\sigma}\left[\sqrt{-g}\left(\tau^{\sigma}_{\phantom{\sigma}{\lambda | f\left(R\right)}}+T^{\sigma}_{\phantom{\sigma}{\lambda}}\right)\right]=0\,.
\end{equation} 
Similar considerations can be developed also in the case of $f(T)$ gravity as we are going to do below.

\section{The gravitational energy-momentum  pseudotensor of $f\left(T\right)$ gravity}\label{GEMPT}
In TEGR and its generalizations, vierbiens are the fields that have to be considered for variations.
Let us take into account a generic analytic $f(T)$ Lagrangian depending on the torsion scalar $T$. The action is: 
\begin{equation}
\mathcal{S}_{f\left(T\right)}=\frac{1}{2\kappa^{2}}\int_{\Omega}d^{4}x  h f\left(T\right)\,,
\end{equation}
where    $h=\text{det}\left(h^{a}_{\phantom{a} {\rho}}\right)$ is the tetrad determinant and, as above,  $\kappa^{2}=8\pi G/c^{4}$. Varying the action  with respect to  $h^{a}_{\ \rho}$ at a  fixed  $x$ and  imposing stationarity, we have:
\begin{gather*}
\delta\mathcal{S}_{f\left(T\right)}=\frac{1}{2\kappa^{2}}\int_{\Omega}d^{4}x\left[hf_{T}\delta T+f\left(T\right)\delta h\right]\\
=\frac{1}{2\kappa^{2}}\int_{\Omega}d^{4}x\Bigl\{\left[4\partial_{\sigma}\left( h f_{T}S_{a}^{\phantom{a}{\rho\sigma}}\right)-4hf_{T}T^{\mu}_{\phantom{\mu}{\nu a}}S_{\mu}^{\phantom{\mu}{\nu\rho}}+f\left(T\right)h h_{a}^{\phantom{a}{\rho}}\right]\delta h^{a}_{\phantom{a}\rho}+\partial_{\sigma}\left(4S_{a}^{\phantom{a}\rho\sigma}hf_{T}\delta h^{a}_{\phantom{a}\rho}\right)\Bigr\}=0\,.
\end{gather*}
Adopting  Eqs.(\ref{formula1}), (\ref{formula2}), (\ref{varh}),  (\ref{varT}) and  imposing the total variation  $\delta h^{a}_{\phantom{a}\rho}\vert_{\partial\Omega}=0$  at the boundary, we obtain the  field equations in vacuum 
\begin{equation}\label{ECFTV}
4h^{-1}\partial_{\sigma}\left[hf_{T}S_{a}^{\phantom{a}{\rho\sigma}}\right]-4f_{T}T^{\mu}_{\phantom{\mu}{\nu a}}S_{\mu}^{\phantom{\mu}{\nu\rho}}+f\left(T\right) h_{a}^{\phantom{a}{\rho}}=0\,.
\end{equation}
 Let us vary now the action  $\mathcal{S}_{f\left(T\right)}$ without  fixing the domain:
\begin{equation}\label{varlocST}
\tilde{\delta}\mathcal{S}_{f\left(T\right)}=\frac{1}{2\kappa^{2}}\int_{\Omega}d^{4}x\left[\delta\left(hf\left(T\right)\right)+\partial_{\mu}\left(hf\left(T\right) \delta x^{\mu}\right)\right]\,,
\end{equation}
we obtain, for a rigid translation \eqref{TrRig} and imposing that the tetrads satisfy \eqref{ECFTV}:
\begin{equation}
\delta h^{a}_{\phantom{a}\rho}=-\epsilon^{\mu}\partial_{\mu}h^{a}_{\phantom{a}\rho}\,,
\end{equation}
\begin{equation}
\tilde{\delta}\mathcal{S}_{f\left(T\right)}=0 \Longleftrightarrow 
\partial_{\sigma}\left(h\tau^{\sigma}_{\phantom{\sigma}{\lambda | f\left(T\right)}}\right)=0\,,
\end{equation}
where
\begin{equation}
\boxed{
2\kappa^{2}\tau^{\sigma}_{\phantom{\sigma}{\lambda | f\left(T\right)}}=-4f_{T}S_{a}^{\phantom{a}{\rho\sigma}}h^{a}_{\phantom{a}{\rho,\lambda}}-f\left(T\right)\delta_{\lambda}^{\sigma}
}
\end{equation}
this is the \emph{the gravitational energy-momentum pseudotensor of $f\left(T\right)$ gravity}. Furthermore, it is  $S_{a}^{\phantom{a}\rho\sigma}=h_{a}^{\phantom{a}\mu}S_{\mu}^{\phantom{\mu}\rho\sigma}$. 
For TEGR, one has:
\begin{equation}
\mathcal{S}_{TEGR}=\frac{1}{2\kappa^{2}}\int_{\Omega}d^{4}x  h T
\end{equation}
and then the gravitational energy-momentum pseudotensor is 
\begin{equation}
\boxed{
\tau^{\sigma}{}_{\lambda |TEGR}=-\frac{2}{\kappa^{2}}S_{\eta}^{\phantom{\eta}{\rho\sigma}}\tilde{\Gamma}^{\eta}_{\phantom{\eta}{\rho\lambda}}-\frac{T}{2\kappa^{2}}\delta_{\lambda}^{\sigma}
}
\end{equation}
where $\tilde{\Gamma}$ is now the   Weitzenb\"ock connection.
In presence of  matter, by varying the action, we have: 
\begin{equation}\label{ECMFT}
2h^{-1}\partial_{\sigma}\left[ h f_{T}S_{a}^{\phantom{a}{\rho\sigma}}\right]-2f_{T}T^{\mu}_{\phantom{\mu}{\nu a}}S_{\mu}^{\phantom{\mu}{\nu\rho}}+\frac{1}{2}f\left(T\right) h_{a}^{\phantom{a}{\rho}}=\kappa^{2}\mathcal{T}^{\left(m\right)\rho}_{a}\,,
\end{equation}
where the energy-momentum tensor of matter fields is 
\begin{equation}
\mathcal{T}^{\left(m\right)\rho}_{a}=-\frac{1}{h}\frac{\delta L_{m}}{\delta h^{a}_{\phantom{a}{\rho}}}\,.
\end{equation}
Using Eqs.\eqref{TENSRICC}  and \eqref{Scalare di curvat}, Eqs.\eqref{ECMFT} assume the form:
\begin{equation}\label{ECT2}
H_{\mu\nu}=f_{T}G_{\mu\nu}+\frac{1}{2}g_{\mu\nu}\left[f\left(T\right)-f_{T}T\right]+2f_{TT}S_{\nu\mu\sigma}\nabla^{\sigma}T=\kappa^{2}\mathcal{T}_{\mu\nu}\,,
\end{equation}
where $G_{\mu\nu}$ is the Einstein tensor and  $\mathcal{T}_{\mu}^{\phantom{\mu}\nu}=h^{a}_{\phantom{a}\mu}\mathcal{T}_{a}^{\phantom{a}\nu}$.  From the divergence of the first term of  \eqref{ECT2}, i.e.  $H_{\mu\nu}$,  using Eqs. \eqref{TENSRICC} and the  anti-symmetry of   contortion $K_{\left(\rho\mu\right)\nu}=0$, we have:
\begin{equation}
\nabla^{\mu}H_{\mu\nu}=-2f_{TT}S^{\rho\sigma}_{\phantom{\rho\sigma}\mu}K_{\sigma\rho\nu}\nabla^{\mu}T=-H_{\lambda\alpha}K^{\lambda\alpha}_{\phantom{\lambda\alpha}\nu}=-H_{\left(\lambda\alpha\right)}K^{\lambda\alpha}_{\phantom{\lambda\alpha}\nu}-H_{\left[\lambda\alpha\right]}K^{\lambda\alpha}_{\phantom{\lambda\alpha}\nu}\,.
\end{equation} 
Considering the symmetry of the energy-momentum tensor  $\mathcal{T}_{\mu}^{\phantom{\mu}\nu}$, the anti-symmetry of the first two indices  of the  contortion, $K_{\left(\rho\mu\right)\nu}=0$, and  $H_{\left[\mu\nu\right]}=0$, we obtain:
\begin{equation}
\nabla^{\mu}H_{\mu\nu}=0\,.
\end{equation}
This means that,  also in teleparallelism, the following   Bianchi relations hold:  
\begin{equation}\label{IBFR2}
\nabla^{\nu}G_{\mu\nu}=0\leftrightarrow\nabla^{\mu}H_{\mu\nu}=0\leftrightarrow\nabla^{\nu}\mathcal{T}^{\left(m\right)}_{\mu\nu}=0\,.
\end{equation}
The local variation of the  $f\left(T\right)$ action for a rigid translation, considering also the matter fields  \eqref{ECMFT},  gives:
\begin{equation}\label{varlocFT1}
0=\frac{h}{\kappa^{2}}\left(P_{f\left(T\right)}\right)_{a}^{\phantom{a}{\rho}}\delta h^{a}_{\phantom{a}\rho}-\partial_{\sigma}\left(h\tau^{\sigma}_{\phantom{\sigma}\lambda}\right)\epsilon^{\lambda}=\left[-h\mathcal{T}_{a}^{\phantom{a}\rho}h^{a}_{\phantom{a}\rho,\lambda}-\partial_{\sigma}\left(h\tau^{\sigma}_{\phantom{\sigma}\lambda}\right)\right]\epsilon^{\lambda}\,,
\end{equation}
where 
\begin{equation}
\left(P_{f\left(T\right)}\right)_{a}^{\phantom{a}{\rho}}=\frac{\kappa^{2}}{h}\frac{\delta L_{f\left(T\right)}}{\delta h^{a}_{\phantom{a}{\rho}}}\,.
\end{equation}
From the symmetry of  $\mathcal{T}_{a}^{\phantom{a}\rho}$:
\begin{equation}
h\nabla_{\eta}\mathcal{T}^{\eta}_{\phantom{\eta}\alpha}=\partial_{\eta}\left(h \mathcal{T}^{\eta}_{\phantom{\eta}\alpha}\right)-h h^{a}_{\phantom{a}\rho,\alpha}\mathcal{T}_{a}^{\phantom{a}\rho}\,,
\end{equation}
and from  \eqref{varlocFT1},  we get:
\begin{equation}
h\nabla_{\eta}\mathcal{T}^{\eta}_{\phantom{\eta}\lambda}-\partial_{\eta}\left(h\mathcal{T}^{\eta}_{\phantom{\eta}\lambda}\right)-\partial_{\sigma}\left(h\tau^{\sigma}_{\phantom{\sigma}\lambda}\right)=0\Rightarrow \partial_{\sigma}\left[h\left(\mathcal{T}^{\sigma}_{\phantom{\sigma}\lambda}+\tau^{\sigma}_{\phantom{\sigma}\lambda}\right)\right]=h\nabla_{\eta}\mathcal{T}^{\eta}_{\phantom{\eta}\lambda}\,.
\end{equation}
Finally, from the Bianchi identities  \eqref{IBFR2}, we obtain the conservation of the matter energy-momentum tensor   and the gravitational energy-momentum pseudotensor, that is:
\begin{equation}\label{consegemt}
\partial_{\sigma}\left[h\left(\tau^{\sigma}_{\phantom{\sigma}{\lambda | f\left(T\right)}}+\mathcal{T}^{\sigma}_{\phantom{\sigma}{\lambda}}\right)\right]=0
\end{equation} 
Eq.\eqref{consegemt}  can be derived also by using the anti-symmetry of the superpotential 
 $S_{\mu}^{\phantom{\mu}\nu\rho}$ after writing the field Eqs.\eqref{ECMFT} in presence of matter  and adopting only the spacetime indices, that is:
\begin{equation}
2h^{-1}\partial_{\sigma}\left[hf_{T}S_{\lambda}^{\phantom{\lambda}\rho\sigma}\right]+2f_{T}\tilde{\Gamma}^{\mu}_{\phantom{\mu}\eta\lambda}S_{\mu}^{\phantom{\mu}\eta\rho}+\frac{1}{2}f\left(T\right)\delta_{\lambda}^{\rho}=\kappa^{2}\mathcal{T}^{\rho}_{\phantom{\rho}\lambda}\,.
\end{equation}
By this approach, we have the  anti-symmetry of superpotential and the commutativity of partial derivatives:
\begin{equation}
\partial_{\rho}\partial_{\sigma}\left[hf_{T}S_{\lambda}^{\phantom{\lambda}\rho\sigma}\right]=-\partial_{\rho}\partial_{\sigma}\left[hf_{T}S_{\lambda}^{\phantom{\lambda}\rho\sigma}\right]=0\,,
\end{equation}
that is 
\begin{equation}
2\partial_{\rho}\partial_{\sigma}\left[hf_{T}S_{\lambda}^{\phantom{\lambda}\rho\sigma}\right]=\partial_{\rho}\left[h\kappa^{2}\mathcal{T}^{\rho}_{\phantom{\rho}\lambda}+h\kappa^{2}\left(-\frac{2}{\kappa^{2}}f_{T}S_{\mu}^{\phantom{\mu}\eta\rho}\tilde{\Gamma}^{\mu}_{\phantom{\mu}\eta\lambda}-\frac{f\left(T\right)}{2\kappa^{2}}\delta_{\lambda}^{\rho}\right)\right]=0\,,
\end{equation}
from which we obtain  the conservation law \eqref{consegemt}. 

The gravitational energy-momentum pseudotensors in metric and tetradic formalisms can be connected considering the role of boundary terms. By these further terms, it is possible to construct a further gravitational pseudotensor by which it is possible to connect the two representations.

\section{The gravitational energy-momentum pseudotensor  for the 
$\omega\left(T,B\right)$ term}\label{GEMPB}
Theories like  $f\left(R\right)$ and  $f\left(T\right)$ are not  equivalent  as GR and   TEGR, that is the linear cases in the curvature $R$ and torsion $T$ scalars (see \cite{CCLS} for a discussion). In fact, while it is possible to formulate  $f\left(R\right)$ gravity by a conformal  transformation  of the  metric as  GR plus a scalar field, this is not possible, in general,   for  $f\left(T\right)$ (see \cite{Bamba}). Furthermore, in a pure metric formalism, $f\left(R\right)$ dynamics is given by fourth-order field equations while $f\left(T\right)$ field equations are second order. Finally,   $f\left(R\right)$ gravity  is invariant under local Lorentz transformations  while   teleparallel $f\left(T\right)$ gravity  depends on the chosen   tetrad frame. However, we can always define a teleparallel equivalent version of  $f\left(R\right)$ gravity considering 
\begin{equation}
f\left(R\left(h\right)\right)=f\left(T\right)+\omega\left(T,B\right)\,,
\end{equation}
where  $T$ is the torsion scalar  and  $B=\left(2/h\right)\partial_{\mu}\left(hT^{\mu}\right)=2\nabla_{\mu}T^{\mu}$ is the boundary term  by which we have $R\left(h\right)=-T-B$ \cite{BBW, BC}.

Let us calculate now the gravitational energy-momentum pseudotensor related to the boundary term $B$ taking into account the action 
 $\mathcal{S}_{B}$ which will allow us to pass from  the gravitational energy-momentum pseudotensor   $\tau^{\alpha}_{\phantom{\alpha}{\lambda\vert f\left(R\right)}}$, coming from  $f\left(R\right)$ gravity, to the gravitational energy-momentum pseudotensor  $\tau_{\phantom{\alpha}\lambda\vert f\left(T\right)}^{\alpha}$, coming from  $f\left(T\right)$ gravity. We start from
\begin{equation}
\mathcal{S}_{B} =\frac{1}{2\kappa^{2}}\int_{\Omega}d^{4}x  h \omega\left(T,B\right)\,.
\end{equation}
Let us vary the action  $\mathcal{S}_{B}$ with respect to the  tetrads  $h^{a}_{\phantom{a}\rho}$ at a fixed  $x$ by using  the variations in   Appendix \ref{B} and  the  relations  (\ref{antdercov}), (\ref{dercovh}),  (\ref{change23}).  We have
\begin{equation}
\delta\mathcal{S}_{B} =\frac{1}{2\kappa^{2}}\int_{\Omega}d^{4}x  \left[\omega\left(T,B\right)\delta h +h\omega_{T}\delta T+h\omega_{B}\delta B\right)]\,,
\end{equation}
and then
\begin{multline}
2\kappa^{2}\delta L_{B}=\Bigl[4\partial_{\sigma}\left(h\omega_{T}S_{a}^{\phantom{a}\rho\sigma}\right)-4h\omega_{T}T^{\mu}_{\phantom{\mu}\nu a}S_{\mu}^{\phantom{\mu}\nu\rho}+\omega hh_{a}^{\phantom{a}\rho}-B\omega_{B}hh_{a}^{\phantom{a}\rho}+2hh_{a}^{\phantom{a}\rho}\Box\omega_{B}\\-2hh_{a}^{\phantom{a}\sigma}\nabla_{\sigma}\nabla^{\rho}\omega_{B}+4h\partial_{\lambda}\omega_{B}S_{a}^{\phantom{a}\lambda\rho}\Bigr]\delta h^{a}_{\phantom{a}\rho}-\partial_{\sigma}\Biggl\{\Bigl[4h\omega_{T}S_{a}^{\phantom{a}\rho\sigma}+2h\partial_{\lambda}\omega_{B}\left(h_{a}^{\phantom{a}\rho}g^{\lambda\sigma}-h_{a}^{\phantom{a}\sigma}g^{\lambda\rho}\right)\\-2\omega_{B}T^{\sigma}hh_{a}^{\phantom{a}\rho}+2\omega_{B}h\left(T^{\rho\sigma}_{\phantom{\rho\sigma}a}+h_{a}^{\phantom{a}\sigma}T^{\rho}+g^{\sigma\rho}T_{a}\right)\Bigr]\delta h^{a}_{\phantom{a}\rho}-2\omega_{B}h\left(h_{a}^{\phantom{a}\rho}g^{\sigma\eta}-h_{a}^{\phantom{a}\eta}g^{\sigma\rho}\right)\delta\left(\partial_{\eta}h^{a}_{\phantom{a}\rho}\right)\Biggr\}\,,
\end{multline}
where $\omega_{T}=\partial\omega/\partial T$, $\omega_{B}=\partial\omega/\partial B$,  and $2\kappa^{2}L_{B}=h\omega\left(T,B\right)$.
Imposing that, at the boundary of the domain, tetrad and first derivative variations,   $\delta h^{a}_{\phantom{a}\rho}$ and  $\delta\left(\partial_{\eta}h^{a}_{\phantom{a}\rho}\right)$,  are zero, we obtain the field equations relative to the boundary terms, that is
\begin{multline}\label{ECBORDIV}
4\partial_{\sigma}\left[ h \omega_{T}S_{a}^{\phantom{a}{ \rho\sigma}}\right]-4h\omega_{T}T^{\mu}_{\phantom{\mu}{ \nu a}}S_{\mu}^{\phantom{\mu}{ \nu\rho}}+\omega\left(T,B\right)h h_{a}^{\phantom{a}{ \rho}}-h\omega_{B}Bh_{a}^{\phantom{a}{\rho}}\\
+2hh_{a}^{\phantom{a}{\rho}}\Box\omega_{B}-2hh_{a}^{\phantom{a}{\sigma}}\nabla_{\sigma}\nabla^{\rho}\omega_{B}+4h\partial_{\mu}\omega_{B}S_{a}^{\phantom{a}{ \mu\rho}}=0\,.
\end{multline}
For a generic  variation of fields and coordinates, it is:
\begin{equation}
\delta L_{B}+\partial_{\mu}\left(L_{B}\delta x^{\mu}\right)=\left(P_{B}\right)_{a}^{\phantom{a}{\rho}}\delta h^{a}_{\phantom{a}\rho}+\partial_{\sigma}\,,\left(hJ^{\sigma}\right)
\end{equation} 
where  $J^{\sigma}$ is the   Noether current and   $\left(P_{\omega}\right)_{a}^{\phantom{a}{\rho}}$ is:
\begin{equation}
\left(P_{B}\right)_{a}^{\phantom{a}{\rho}}=\frac{1}{h}\frac{\delta L_{B}}{\delta h^{a}_{\phantom{a}{\rho}}}\,.
\end{equation}
Then we have
\begin{multline}
-2\kappa^{2}J^{\sigma}=\Bigl[4\omega_{T}S_{a}^{\phantom{a}\rho\sigma}+2\partial_{\lambda}\omega_{B}\left(h_{a}^{\phantom{a}\rho}g^{\lambda\sigma}-h_{a}^{\phantom{a}\sigma}g^{\lambda\rho}\right)-2\omega_{B}T^{\sigma}h_{a}^{\phantom{a}\rho}\\+2\omega_{B}\left(T^{\rho\sigma}_{\phantom{\rho\sigma}a}+h_{a}^{\phantom{a}\sigma}T^{\rho}+g^{\sigma\rho}T_{a}\right)\Bigr]\delta h^{a}_{\phantom{a}\rho}-2\omega_{B}\left(h_{a}^{\phantom{a}\rho}g^{\sigma\eta}-h_{a}^{\phantom{a}\eta}g^{\sigma\rho}\right)\delta\left(\partial_{\eta}h^{a}_{\phantom{a}\rho}\right)-\omega \delta x^{\sigma}\,.
\end{multline}
If the action is invariant under rigid translations and tetrads verify the field equations  \eqref{ECBORDIV}, we have,  assuming  $\delta \left(\partial_{\eta}h^{a}_{\phantom{a}\rho}\right)=-h^{a}_{\phantom{a}\rho,\lambda\eta}\epsilon^{\lambda}$, for  translations
\begin{equation}
\tilde{\delta}\mathcal{S}_{B}=0 \Rightarrow \partial_{\sigma}\left(h \tau_{\phantom{\sigma}\mu |\omega}^{\sigma}\right)=0\,,
\end{equation}
where $J^{\sigma}=-\tau_{\phantom{\sigma}\lambda |\omega}^{\sigma}\epsilon^{\lambda}$:
\begin{equation}\label{EMTGB}
\boxed{
\begin{split}
2\kappa^{2}\tau_{\phantom{\sigma}\lambda\vert\omega\left(T,B\right)}^{\sigma}=&-\Bigl[4\omega_{T}S_{a}^{\phantom{a}\rho\sigma}+2\omega_{B}\left(T^{\rho\sigma}_{\phantom{\rho\sigma}a}+h_{a}^{\phantom{a}\sigma}T^{\rho}+g^{\sigma\rho}T_{a}-T^{\sigma}h_{a}^{\phantom{a}\rho}\right)\\&+2\partial_{\eta}\omega_{B}\left(h_{a}^{\phantom{a}\rho}g^{\eta\sigma}-h_{a}^{\phantom{a}\sigma}g^{\eta\rho}\right)\Bigr]h^{a}_{\phantom{a}\rho,\lambda}+2\omega_{B}\left(h_{a}^{\phantom{a}\rho}g^{\sigma\eta}-h_{a}^{\phantom{a}\eta}g^{\sigma\rho}\right)h^{a}_{\phantom{a}\rho,\lambda\eta}-\omega\delta^{\sigma}_{\lambda}
\end{split}
}
\end{equation}
with  $\tau_{\phantom{\sigma}\lambda\vert\omega}^{\sigma}$ the  \emph{gravitational energy-momentum pseudotensor of $\omega\left(T,B\right)$}. This object is  due to the boundary terms.
The calculation of  psudotensor $\tau_{\phantom{\sigma}\lambda |\omega}^{\sigma}$ could be performed directly  from  derivatives  instead of using  the variations. Let us express the boundary term  $B$ in terms of  tetrads $h_{a}^{\phantom{a}{\rho}}$ and its derivatives:
\begin{multline}
B=4\Bigl[h_{a}^{\phantom{a}[\rho}g^{\sigma]\mu}\partial_{\mu}\partial_{\sigma}h^{a}_{\phantom{a}\rho}+\bigl(h_{a}^{\phantom{a}\rho}h_{b}^{\phantom{b}[\tau}g^{\lambda]\mu}-h_{a}^{\phantom{a}[\lambda}h_{b}^{\phantom{b}\tau]}g^{\mu\rho}-h_{a}^{\phantom{a}\mu}h_{b}^{\phantom{b}[\tau}g^{\lambda]\rho}\\
-h_{a}^{\phantom{a}\tau}h_{b}^{\phantom{b}[\rho}g^{\mu]\lambda}\bigr)\partial_{\mu}h^{a}_{\phantom{a}\rho}\partial_{\lambda}h^{b}_{\phantom{b}\tau}\Bigr]\,.
\end{multline}
Using the  gravitational  energy-momentum pseudotensor for a generic Lagrangian depending up to  second order tetrad derivatives  (see  \cite{CCT}) and adopting the formulas  (\ref{formula3}), (\ref{formula4}), (\ref{formula5}), i.e.
\begin{equation}
h\tau^{\sigma}_{\phantom{\sigma}\mu}=\left(\frac{\partial L_{B}}{\partial\partial_{\sigma}h^{a}_{\phantom{a}\rho}}-\partial_{\lambda}\frac{\partial L_{B}}{\partial\partial_{\lambda\sigma}h^{a}_{\phantom{a}\rho}}\right)h^{a}_{\phantom{a}{ \rho,\mu}}+\frac{\partial L_{B}}{\partial\partial_{\lambda\sigma}h^{a}_{\phantom{a}\rho}}h^{a}_{\phantom{a}\rho,\mu\lambda}-\delta^{\sigma}_{\mu}L_{B}\,,
\end{equation}
we have
\begin{multline}
2\kappa^{2}\tau_{\phantom{\sigma}\mu\vert\omega}^{\sigma}=\bigl[\omega_{T}\frac{\partial T}{\partial\partial_{\sigma}h^{a}_{\phantom{a}\rho}}+\omega_{B}\frac{\partial B}{\partial\partial_{\sigma}h^{a}_{\phantom{a}{ \rho}}}-h^{-1}\partial_{\lambda}\bigl(h\omega_{B}\frac{\partial B}{\partial\partial_{\lambda\sigma}h^{a}_{\phantom{a}{ \rho}}}\bigr)\bigr]h^{a}_{\phantom{a}{ \rho,\mu}}\\
+\omega_{B}\frac{\partial B}{\partial\partial_{\lambda\sigma}h^{a}_{\phantom{a}{ \rho}}}h^{a}_{\phantom{a}\rho,\mu\lambda}-\omega\left(T,B\right)\delta_{\mu}^{\sigma}\,.
\end{multline}
Considering the following derivatives of $B$ and $T$, one has  
\begin{multline}
\frac{\partial B}{\partial\partial_{\sigma}h^{a}_{\phantom{a} {\rho}}}=2\left(h_{a}^{\phantom{a}\rho}T^{\sigma}-T_{a}g^{\sigma\rho}-h_{a}^{\phantom{a}\sigma}T^{\rho}-T^{\rho\sigma}_{\phantom{\rho\sigma}a}\right)+4\Bigl(h_{a}^{\phantom{a}\tau}h_{b}^{\phantom{b}[\sigma}g^{\rho]\lambda}+h_{a}^{\phantom{a}[\rho\vert}h_{b}^{\phantom{b}\tau}g^{\vert\sigma]\lambda}\\
-h_{a}^{\phantom{a}[\rho}h_{b}^{\phantom{b}\sigma]}g^{\tau\lambda}-h_{a}^{\phantom{a}[\rho\vert}h_{b}^{\phantom{b}\lambda}g^{\vert\sigma]\tau}\Bigr)\partial_{\lambda}h^{b}_{\phantom{b}\tau}\,,
\end{multline}
\begin{equation}
\frac{\partial B}{\partial\partial_{\lambda\sigma}h^{a}_{\phantom{a}{\rho}}}=2\left(h_{a}^{\phantom{a} {\rho}}g^{\lambda\sigma}-h_{a}^{\phantom{a}{\sigma}}g^{\lambda\rho}\right)=4h_{a}^{\phantom{a}[\rho}g^{\sigma]\lambda}\,,
\end{equation}
and 
\begin{equation}
\frac{\partial T}{\partial\partial_{\sigma}h^{a}_{\phantom{a}\rho}}=-4S_{a}^{\phantom{a}\rho\sigma}\,.
\end{equation}
Combining these results, it is straightforward to obtain the expression \eqref{EMTGB}.

\section{ From $f\left(R\right)$ gravity to $f\left(T\right)$ gravity and viceversa by $\tau_{\lambda |\omega\left(T,B\right)}^{\ \sigma}$}\label{JUMPTB}

Also if the two theories of gravity are not equivalent, it is possible to pass from $f(R)$  to $f(T)$ (and viceversa) by means of the pseudotensor  
 $\tau_{\phantom{\sigma}\lambda\vert\omega\left(T,B\right)}^{\sigma}$ being:
\begin{equation}\label{RTOMEGA}
\boxed{
\tau_{\phantom{\sigma}\lambda\vert f\left(R\right)}^{\sigma}=\tau_{\phantom{\sigma}\lambda\vert f\left(T\right)}^{\sigma}+\tau_{\phantom{\sigma}\lambda\vert\omega\left(T,B\right)}^{\sigma}
}
\end{equation}
The demonstration of this statement can be easily achieved as follows.
Let us   locally vary  the action for  $f\left(R\right)$ with  respect  to the  metric $g_{\mu\nu}$ and the two actions  for  $f\left(T\right)$ and  $\omega\left(T,R\right)$ with respect to the tetrads   $h^{a}_{\phantom{a}\rho}$,  we have:
 \begin{equation}
 \begin{split}
 \tilde{\delta}_{g,x}\mathcal{S}_{f\left(R\right)}&=\frac{1}{2\kappa^{2}}\int_{\Omega}d^{4}x\left[\sqrt{-g}P_{f\left(R\right)}^{\mu\nu}\delta g_{\mu\nu}+\partial_{\lambda}\left(2\kappa^{2}\sqrt{-g}J_{f\left(R\right)}^{\lambda}\right)\right]\,,\\
 \tilde{\delta}_{h,x}\mathcal{S}_{f\left(T\right)}&=\frac{1}{2\kappa^{2}}\int_{\Omega}d^{4}x\left[2h\left(P_{f\left(T\right)}\right)_{a}^{\phantom{a} {\rho}}\delta h^{a}_{\phantom{a}{\rho}}+\partial_{\lambda}\left(2\kappa^{2}hJ_{f\left(T\right)}^{\lambda}\right)\right]\,,\\
\tilde{\delta}_{h,x}\mathcal{S}_{\omega\left(T,B\right)}&=\frac{1}{2\kappa^{2}}\int_{\Omega}d^{4}x\left[2h\left(P_{\omega}\right)_{a}^{\phantom{a}{\rho}}\delta h^{a}_{\phantom{a}{\rho}}+\partial_{\lambda}\left(2\kappa^{2}hJ_{\omega\left(T,B\right)}^{\lambda}\right)\right]\,, 
 \end{split}
 \end{equation}
 where  $J^{\lambda}$ are  the  Noether currents.
 From
 \begin{equation}
 \mathcal{S}_{f\left(R\right)}\left(h\right)=\mathcal{S}_{f\left(T\right)}+\mathcal{S}_{\omega\left(T,B\right)}\,,
 \end{equation}
 we have
 \begin{equation}
\tilde{\delta}_{g,x}\mathcal{S}_{f\left(R\right)}= \tilde{\delta}_{h,x}\mathcal{S}_{f\left(T\right)}+\tilde{\delta}_{h,x}\mathcal{S}_{\omega\left(T,B\right)}\,,
\end{equation}
that is 
\begin{equation}
\begin{split}
\sqrt{-g}P_{f\left(R\right)}^{\mu\nu}\delta g_{\mu\nu}+\partial_{\lambda}\left(2\kappa^{2}\sqrt{-g}J_{f\left(R\right)}^{\lambda}\right)=2h\left[\left(P_{f\left(T\right)}\right)_{a}^{\phantom{a}{\rho}}+\left(P_{\omega}\right)_{a}^{\phantom{a}{\rho}}\right]\delta h^{a}_{\phantom{a}{\rho}}\\
+\partial_{\lambda}\left[2\kappa^{2}h\left(J_{f\left(T\right)}^{\lambda}+J_{\omega\left(T,B\right)}^{\lambda}\right)\right]\,.
\end{split}
\end{equation}
Since the following identities hold by expressing the r.h.s. in tetrad terms
\begin{equation}
\sqrt{-g}P_{f\left(R\right)}^{\mu\nu}\delta g_{\mu\nu}\vert[h]=2h\left[\left(P_{f\left(T\right)}\right)_{a}^{\phantom{a}{\rho}}+\left(P_{\omega}\right)_{a}^{\phantom{a}{\rho}}\right]\delta h^{a}_{\phantom{a}{\rho}}\,,
\end{equation}
we have
\begin{equation}
\partial_{\lambda}\left(\sqrt{-g}J_{f\left(R\right)}^{\lambda}\right)=\partial_{\lambda}\left[h\left(J_{f\left(T\right)}^{\lambda}+J_{\omega\left(T,B\right)}^{\lambda}\right)\right]\,,
\end{equation}
from which  Eq.\eqref{RTOMEGA} holds for  rigid translations thanks to the Noether currents. In other words,  by this  formalism based on pseudotensors,  it is possible to relate metric and tetradic pictures of $f(R)$ and $f(T)$ theories of gravity.

\section{Conclusions}\label{CONC}
The gravitational energy-momentum pseudotensor $\tau^{\alpha}_{\phantom{\alpha}\lambda}$  is  an important feature of gravitational interaction  capable of discriminating among   theories of gravity.  In this paper, we discussed  this pseudotensor comparing its derivation in GR,  in TEGR,  and in their straightforward generalizations, namely $f(R)$  and $f(T)$ gravity.  The considerations have been developed in  both  metric and  vierbien formalisms. The main result is that the two pseudotensors are related defining a further pseudotensor,
$\tau_{\phantom{\sigma}\lambda\vert\omega\left(T,B\right)}^{\sigma}$, derived from the boundary term $B$ that connects the curvature  and the torsion scalars as $R=-T-B$. This characteristic could assume a relevant role in tests of gravity aimed to put in evidence further terms and degrees of freedom with respect to GR. In particular, as discussed in detail in \cite{Felix}, the gravitational pseudotensor  is important to point out differences in quadrupolar gravitational radiations coming from GR and $f(R)$ gravity. Using a similar approach in metric and teleparallel gravity, discrimination of theories can be achieved considering gravitational radiation:  further  polarizations and energy-momentum contributions can be selected and classified in view of possible observations. For example, as discussed in \cite{Bamba1,Abedi}, gravitational waves in $f(T)$ gravity and its generalizations are substantially different with respect to gravitational waves in $f(R)$ gravity so detecting further polarization modes of gravitational radiation could be a way to discriminate among theories. The issue of degrees of freedom in $f(T)$ gravity is discussed in detail in \cite{guzman}.  

In a forthcoming paper, we will discuss the weak field limit of   gravitational pseudotensor in the various formulations in view of selecting physical observables and discussing possible experimental limits on gravitational radiation.

\section*{Acknowledgements}
SC acknowledges INFN Sez. di Napoli (Iniziative Specifiche QGSKY and TEONGRAV).  
This article is based upon work from COST Action (CA15117, CANTATA), supported by COST (European Cooperation in Science and Technology).

\appendix 
\section{Appendix: Useful formulae}\label{A}
\begin{equation}\label{trinabla}
\nabla^{\nu}\nabla_{\mu}\nabla_{\nu} f\left(R\right)=R^{\alpha}_{\phantom{\alpha}\mu}\nabla_{\alpha}f\left(R\right)+\nabla_{\mu}\Box f\left(R\right)
\end{equation}
\begin{equation}\label{formula1}
\frac{\partial h _{c}^{\phantom{c}\nu}}{\partial h^{a}_{\phantom{a}\rho}}=-h_{a}^{\phantom{a}\nu}h_{c}^{\phantom{c}\rho}
\end{equation}
\begin{equation}\label{formula2}
\frac{\partial g^{\mu\nu}}{\partial h^{c}_{\phantom{c}\rho}}=-g^{\rho\nu}h_{c}^{\phantom{c}\mu}-g^{\rho\mu}h_{c}^{\phantom{c}\nu}
\end{equation}
\begin{equation}\label{formula3}
\partial_{\alpha}h_{a}^{\phantom{a}\rho}=-h_{a}^{\phantom{a}\mu}h_{b}^{\phantom{b}\rho}\partial_{\alpha}h^{b}_{\phantom{b}\mu}=-\tilde{\Gamma}^{\rho}_{\phantom{\rho}\lambda\alpha}h_{a}^{\lambda}
\end{equation}
\begin{equation}\label{formula4}
\frac{\partial h}{\partial h^{a}_{\phantom{a}\rho}}=h_{a}^{\phantom{a}\rho}h
\end{equation}
\begin{equation}\label{formula5}
\partial_{\mu} h=h\tilde{\Gamma}^{\nu}_{\phantom{\nu}\mu\nu}-hK^{\nu}_{\phantom{\nu}\mu\nu}=h\tilde{\Gamma}^{\nu}_{\phantom{\nu}\nu\mu}=hh_{a}^{\phantom{a}\rho}h^{a}_{\phantom{a}\rho,\mu}
\end{equation}
\begin{equation}\label{antdercov}
\partial_{\sigma}\left(hF^{\rho\sigma}\right)=h\nabla_{\sigma}F^{\rho\sigma}\quad\text{se}\quad F^{\left(\rho\sigma\right)}=0 
\end{equation}
\begin{equation}\label{dercovh}
\nabla_{\sigma}h_{a}^{\phantom{a}\rho}=-h_{a}^{\phantom{a}\eta}K^{\rho}_{\phantom{\rho}\eta\sigma}\quad \nabla_{\sigma}h_{a}^{\phantom{a}\sigma}=T_{a}
\end{equation}
\begin{equation}\label{change23}
K^{\rho\phantom{a}\lambda}_{\phantom{\rho}a}=K^{\rho\lambda}_{\phantom{\rho\lambda}a}+T^{\rho\lambda}_{\phantom{\rho\lambda}a}
\end{equation}
\section{Appendix: Variations}\label{B}
\begin{equation}
\delta h_{a}^{\phantom{a}\rho}=-h_{a}^{\phantom{a}\mu}h_{b}^{\phantom{b}\rho}\delta h^{b}_{\phantom{b}\mu}
\end{equation}
\begin{equation}\label{varh}
\delta h=hh_{a}^{\phantom{a}\rho}\delta h^{a}_{\phantom{a}\rho}=-hh^{a}_{\phantom{a}\rho}\delta h_{a}^{\phantom{a}\rho}
\end{equation}
\begin{equation}\label{varT}
\delta T=-4 T^{\mu}_{\phantom{\mu}{\nu a}}S_{\mu}^{\phantom{\mu}{\nu\rho}}\delta h^{a}_{\phantom{a}{\rho}}-4S_{a}^{\phantom{a}{\rho\sigma}}\delta\left(\partial_{\sigma} h^{a}_{\phantom{a}{\rho}}\right)
\end{equation}
\begin{equation}\label{varTLAM}
\delta T^{\lambda}=-\left(T^{\rho\lambda}_{\phantom{\rho\lambda}{a}}+h_{a}^{\phantom{a}{\lambda}}T^{\rho}+g^{\lambda\rho}T_{a}\right)\delta h^{a}_{\phantom{a}{\rho}}+\left(h_{a}^{\phantom{a}{\rho}}g^{\lambda\sigma}-h_{a}^{\phantom{a}{\sigma}}g^{\lambda\rho}\right)\delta\left(\partial_{\sigma} h^{a}_{\phantom{a}{\rho}}\right)
\end{equation}
\begin{equation}\label{varB}
\delta B = \delta \left[\frac{2}{h}\partial_{\sigma}\left(h T^{\nu\sigma}_{\phantom{\nu\sigma}{\nu}}\right)\right]=-\frac{B}{h}\delta h+\frac{2}{h}\partial_{\mu}\left[T^{\mu}\delta h+h\delta T^{\mu}\right]
\end{equation}
\begin{equation}\label{omegavart}
h\omega_{T}\delta T=\left[-4h\omega_{T}T^{\mu}_{\phantom{\mu}\nu a}S_{\mu}^{\phantom{\mu}\nu\rho}+4\partial_{\sigma}\left(h\omega_{T}S_{a}^{\phantom{a}\rho\sigma}\right)\right]\delta h^{a}_{\phantom{a}\rho}-\partial_{\sigma}\left(4h\omega_{T}S_{a}^{\phantom{a}\rho\sigma}\delta h^{a}_{\phantom{a}\rho}\right)
\end{equation}
\begin{equation}\label{omegavarb}
\begin{split}
h\omega_{B}\delta B&=\left[-B\omega_{B}hh_{a}^{\phantom{a}\rho}+2hh_{a}^{\phantom{a}\rho}\Box\omega_{B}-2hh_{a}^{\phantom{a}\sigma}\nabla_{\sigma}\nabla^{\rho}\omega_{B}+4h\partial_{\lambda}\omega_{B}S_{a}^{\phantom{a}\lambda\rho}\right]\delta h^{a}_{\phantom{a}\rho}\\
&-\partial_{\sigma}\left[2h\partial_{\lambda}\omega_{B}\left(h_{a}^{\phantom{a}\rho}g^{\lambda\sigma}-h_{a}^{\phantom{a}\sigma}g^{\lambda\rho}\right)\delta h^{a}_{\phantom{a}\rho}-2\omega_{B}\left(T^{\sigma}\delta h+h\delta T^{\sigma}\right)\right]
\end{split}
\end{equation}


\begin{thebibliography}{99}
\bibitem{CL} S. Capozziello and M. De Laurentis, \emph{Extended Theories of Gravity}, Phys. Rept. \textbf{509}, 167 (2011)
\bibitem{AE}A. Einstein, \emph{Auf die Riemann-Metrik und den Fern-Parallelismus gegr\"undete einheitliche Feldtheorie}, Math. Annal. \textbf{102}, 685 (1930). 
\bibitem{UC} A. Unzicker, T. Case, \emph{Unied Field Theory
based on Riemannian Metrics and distant Parallelism}, arXiv:physics/0503046v1
\bibitem{WIT}R. Weitzenb\"ock, \emph{Invarianten Theorie}, Nordhoff, Groningen (1923)
\bibitem{APTG} R. Aldrovandi and J. G. Pereira, \emph{Teleparallel Gravity: An Introduction}, Fundamental Theories of Physics, vol. \textbf{173} (Springer, Dordrecht,
2013)
\bibitem{Laur}
  M.~Hohmann, L.~Järv, M.~Krssak and C.~Pfeifer,
  \emph{Teleparallel theories of gravity as analogue of non-linear electrodynamics,}
  arXiv:1711.09930 [gr-qc] (2017).
 \bibitem{MAL}J. W. Maluf, \emph{The teleparallel equivalent of general relativity}, Ann. Phys. \textbf{525}, 339 (2013)
\bibitem{CCLS} Y. Cai, S. Capozziello, M. De Laurentis and E. N. Saridakis,  \emph{f(T) teleparallel gravity and cosmology}, Rept. Prog. Phys. \textbf{79}, 106901 (2016)
\bibitem{FFPL}R. Ferraro and F. Fiorini, \emph{Non-trivial frames for f(T) theories of gravity and beyond}, Phys. Lett. B \textbf{702}, 75 (2011)
\bibitem{Bamba} 
  K.~Bamba, S.~D.~Odintsov and D.~Sáez-Gómez,
  \emph{Conformal symmetry and accelerating cosmology in teleparallel gravity},
  Phys.\ Rev.\ D {\bf 88} (2013) 084042
  \bibitem{Emmanuel1}
  M.~Hohmann, L.~Järv and U.~Ualikhanova,
  \emph{Covariant formulation of scalar-torsion gravity,}
  arXiv:1801.05786 [gr-qc] (2018).
\bibitem{Martin1}
  M.~Krssak and E.~N.~Saridakis,
  \emph{The covariant formulation of f(T) gravity,}
  Class.\ Quant.\ Grav.\  {\bf 33},    115009 (2016)
  \bibitem{Martin2}
  M.~Krssak and J.~G.~Pereira,
  \emph{Spin Connection and Renormalization of Teleparallel Action},
  Eur.\ Phys.\ J.\ C {\bf 75},  519 (2015)
  \bibitem{BSB}B. Li, T. P. Sotiriou and J. D. Barrow, \emph{f(T) gravity and local Lorentz invariance}, Phys. Rev. D \textbf{83}, 064035 (2011)
\bibitem{CRTEL}L. Combi, G. E. Romero, \emph{Is Teleparallel Gravity Really Equivalent to General Relativity?}, Ann. Phys., \textbf{530}, 1700175 (2018)
\bibitem{KIB}T. W. B. Kibble, \emph{Lorentz Invariance and the Gravitational Field}, J. Math. Phys. \textbf{2}, 212 (1961) 
\bibitem{FF}R. Ferraro and F. Fiorini, \emph{Remnant group of local Lorentz transformation in f(T) theories}, Phys. Rev. D, \textbf{91}, 064019 (2015)
\bibitem{SLB}T. P. Sotiriou, B. Li and J. D. Barrow, \emph{Generalizations of teleparallel gravity and local Lorentz symmetry}, Phys. Rev. D \textbf{83}, 104030 (2011)
\bibitem{Emmanuel2}
  M.~Hohmann, L.~Jarv and U.~Ualikhanova,
  \emph{Dynamical systems approach and generic properties of $f(T)$ cosmology},
  Phys.\ Rev.\ D {\bf 96} (2017)   043508 (2017).   
 \bibitem{BBW}S. Bahamonde, C. G. Boehmer and M. Wright,
\emph{Modificated teleparallel theories of gravity}, Phys.
Rev. D \textbf{92} 104042 (2015) 
\bibitem{BC}S. Bahamonde, S. Capozziello, \emph{Noether symmetry approach in f (T, B) teleparallel cosmology}, Eur. Phys. J. C
\bibitem{Cho}Y. M. Cho, \emph{Einstein Lagrangian as the translational Yang-Mills Lagrangian}, Phys. Rev. D \textbf{14}, 2521 (1976)
\bibitem{OS} G. Otalora and E. N. Saridakis, \emph{Modified teleparallel gravity with higher-derivative torsion terms}, Phys. Rev. D \textbf{94}, 084021 (2016)
\bibitem{KS}G. Kofinas and E. N. Saridakis, \emph{Teleparallel equivalent of Gauss-Bonnet gravity and its modifications},  Phys. Rev. D \textbf{90}, 084044 (2014)
\bibitem{Kostas}
  S.~Capozziello, M.~De Laurentis and K.~F.~Dialektopoulos,
  \emph{Noether symmetries in Gauss-Bonnet-teleparallel cosmology,}
  Eur.\ Phys.\ J.\ C {\bf 76},  629 (2016)
 \bibitem{felixgauss}
  M.~De Laurentis and A.~J.~Lopez-Revelles,
 \emph{Newtonian, Post Newtonian and Parameterized Post Newtonian limits of $f(R, G)$ gravity},
  Int.\ J.\ Geom.\ Meth.\ Mod.\ Phys.\  {\bf 11} (2014) 1450082  
\bibitem{LLF}L. D. Landau and E. M. Lifshitz, \emph{The Classical Theory of Fields}, Pergamon Press, Oxford (1971)
\bibitem{MWH}F. I. Mikhail, M. I. Wanas, A. Hindawi and E. I. Lashin, \emph{Energy-Momentum Complex in M\o ller's
Tetrad Theory Of Gravitation}, Int. J. Theor. Phys. \textbf{32}, 1627 (1993), https://doi.org/10.1007/BF00672861
\bibitem{CCT}S. Capozziello, M. Capriolo and M. Transirico, \emph{The gravitational energy-momentum  pseudotensor of higher order theories of gravity}, Ann. Phys. \textbf{525}, 1600376 (2017) 
\bibitem{vagenas}
T. Multamaki, A. Putaja, I. Vilja, and E. C. Vagenas, 
\emph{Energy-momentum complexes in $f(R)$ theories of gravity}
 Class.Quant.Grav. {\bf 25} (2008) 075017
 \bibitem{abedi1}
 H. Abedi and M. Salti,
 \emph{Multiple field modified gravity and localized energy in teleparallel framework},
 Gen.Rel. Grav. 47 (2015) 93.
\bibitem{Ali}
  S.~A.~Ali and S.~Capozziello,
  \emph{Nonlinear realization of the local conform-affine symmetry group for gravity in the composite fiber bundle formalism,}
  Int.\ J.\ Geom.\ Meth.\ Mod.\ Phys.\  {\bf 4} (2007) 1041
 \bibitem{KN}S. Kobayashi and K. Nomizu, \emph{Foundation of Differential Geometry}, Vol. I, Wiley Interscience, New York (1963)
\bibitem{LEE}J. M. Lee, \emph{Introduction to Smooth Manifolds}, GTM 218, Springer-Verlag, New York, (2013)
\bibitem{LEER}J. M. Lee, \emph{Riemannian Manifolds}, GTM 176, Springer-Verlag, New York, (1997)
\bibitem{MI}P. M. Michor, \emph{Topics in Differential Geometry}, American Mathematical Society, Providence, RI (2008)
\bibitem{AT} L.P. Eisenhart, \emph{An introduction to differential geometry}, Princeton Univ. Press, Princeton (1947)
\bibitem{NAK}M. Nakahara, \emph{Geometry, topology and physics}, Institute of physics publishing, Bristol and Philadelphia, (2003)
\bibitem{SGMMP}B. Schutz, \emph{Geometrical methods of mathematical physics}, Cambridge University Press, Cambridge (1980)
\bibitem{YCB}Y. Choquet-Bruhat, \emph{General Relativity and the Einstein's Equation}, Oxford University Press, New York (2009)
\bibitem{CR}S. M. Carroll, \emph{Spacetime And Geometry}, Addison Wesley, San Francisco (2004)
\bibitem{WGR}R. M. Wald, \emph{General Relativity}, The University of Chicago Press, Chicago (1984)
\bibitem{GRG}M. Gasperini, \emph{Theory of Gravitational Interactions}, Springer, Dordrecht  (2013)
\bibitem{HS}K. Hayashi and T. Shirafuji, \emph{New General Relativity}, Phys. Rev. D \textbf{19}, 3524 (1979)
\bibitem{AP}R. Aldrovandi and J. G. Pereira, \emph{An Introduction to Geometrical Physics}, World Scientifics, Singapore (1995)
 \textbf{77}, 107 (2017) 
\bibitem{SN}N. Straumann, \emph{General Relativity}, Springer, Dordrecht (2004)
\bibitem{CF}S. Capozziello and V. Faraoni, \emph{Beyond Einstein Gravity}, FTP 170, Springer, New York (2011)
\bibitem{HE}S. W. Hawking and G.F.R. Ellis, \emph{The Large Scale Structure of Spacetime}, Cambridge University Press, Cambridge (1973)
\bibitem{SGMM}B. Schutz, \emph{A First Course in General Relativity}, Cambridge University Press, Cambridge (2009)
\bibitem{OR}H. C. Ohanian and R. Ruffini, \emph{Gravitation and Spacetime}, Cambridge University Press, Cambridge (2013)
\bibitem{BR} H. Stephani, \emph{General Relativity}, Cambridge University Press, Cambridge (1990)
 \bibitem{Felix}
  M.~De Laurentis and S.~Capozziello,
  \emph{Quadrupolar gravitational radiation as a test-bed for $f(R)$ gravity,}
  Astropart.\ Phys.\  {\bf 35}, 257  (2011) 
  \bibitem{Bamba1}
  K.~Bamba, S.~Capozziello, M.~De Laurentis, S.~Nojiri and D.~Sáez-Gómez,
  \emph{No further gravitational wave modes in $F(T)$ gravity},
  Phys.\ Lett.\ B {\bf 727}, 194 (2013)
  \bibitem{Abedi}
  H.~Abedi and S.~Capozziello,
  \emph{Gravitational waves in modified teleparallel theories of gravity,},
  arXiv:1712.05933 [gr-qc] (2017) to be published in EPJC.
  \bibitem{guzman}
  R. Ferraro and M.J Guzman, 
   \emph{Hamiltonian formalism for $f(T)$ gravity}
     Phys. Rev. D {\bf 97}, (2018) 104028   
 
  
  


\end{thebibliography}
\end{document}